\documentclass[runningheads]{llncs}

\usepackage[T1]{fontenc}
\usepackage{graphicx}
\usepackage{amsmath,amssymb}
\usepackage{booktabs}
\usepackage{placeins}
\usepackage{float}
\usepackage{array}
\usepackage{multirow}
\usepackage{algorithm}
\usepackage{algpseudocode}
\usepackage{textcomp}
\usepackage{tabularx}
\usepackage{xcolor}
\usepackage{longtable}
\usepackage{capt-of}
\usepackage{tikz}
\usepackage{url}
\usepackage[colorlinks=true,linkcolor=black,citecolor=black,urlcolor=blue]{hyperref}
\usetikzlibrary{arrows.meta,positioning,fit,calc,shapes.geometric,shapes.misc}

\graphicspath{{figures/}}
\newcolumntype{Y}{>{\raggedright\arraybackslash}X}
\newcolumntype{C}{>{\centering\arraybackslash}X}

\begin{document}

\title{Open-Source Autonomous Driving System Analysis and Multi-Disciplinary Hardware-in-the-Loop Research Paradigm with Reinforcement-Learning Testing and Large Language Models}
\titlerunning{Apollo-on-Hongqi EV Hardware-in-the-Loop Research Paradigm}

\author{Dianjing Cheng\inst{1} \and
Yike Li\inst{1} \and
Lan Yang\inst{6} \and
Shan Fang\inst{6} \and
Wenjia Niu\inst{1}\thanks{Corresponding author.} \and
Xiangyu Shi\inst{1} \and
Xinyi Zhao\inst{1} \and
Yunzhe Tian\inst{1} \and
XingYu Wu\inst{1} \and
Xiaoshu Cui\inst{1} \and
Yuanwan Chen\inst{1} \and
Jialu Sun\inst{1} \and
Zhongli Wang\inst{2} \and
Biao Liu\inst{3} \and
Jiaqi Yang\inst{3} \and
Jinghui Feng\inst{4} \and
Feifei Su\inst{4} \and
Juan Du\inst{4} \and
Shuangde Fang\inst{4} \and
Yi Qian\inst{4} \and
Huiyun Li\inst{8} \and
Yuansheng Liu\inst{7} \and
Peng Sun\inst{4} \and
Mingming Wan\inst{5} \and
Nan Chen\inst{5} \and
Ruipeng Gao\inst{1}}

\authorrunning{D. Cheng et al.}

\institute{School of Cyberspace Science and Technology, Beijing Jiaotong University, Beijing, China
\and
School of Automation and Intelligence, Beijing Jiaotong University, Beijing, China
\and
School of Electrical Engineering, Beijing Jiaotong University, Beijing, China
\and
Data Security Department, Baidu, Inc. Beijing, China
\and
Geely Automobile Research Institute, Zhejiang, China
\and
School of Information Engineering, Chang'an University, Xi'an, China
\and
College of Robotics, Beijing Union University, Beijing, China
\and
Faculty of Computility Microelectronics, Shenzhen University of Advanced Technology, Shenzhen, China
\\
\email{niuwj@bjtu.edu.cn}}

\maketitle

\begin{abstract}
Open-source autonomous driving systems provide an inspectable software foundation for intelligent vehicle research. Under real-vehicle deployment conditions, the recording and review of experimental conditions are important for interpreting system behavior and reusing experimental results. However, in a shared real-vehicle environment involving multiple vehicles, task processes, code modifications, and hardware testing feedback are often distributed across different teams and experimental stages, making it challenging to maintain continuous and reviewable experimental records. To address this limitation, this paper examines an Apollo-on-Hongqi EV environment and proposes a real-vehicle experimental framework. The framework connects multi-vehicle experiments, repository-based code reuse and software-hardware testing feedback within a unified review process. Large language models and RL-based testing serve as auxiliary components for record organization, anomaly summarization, and simulation-based candidate scenario generation. Based on this setting, this paper analyzes preliminary evidence from multi-vehicle collaborative experimentation, code and experimental-skill sharing, and software-hardware collaborative testing. The analysis shows that experimental records can be examined together with their operating conditions, providing a reviewable basis for Apollo-on-Hongqi EV research.

\end{abstract}

\section{Introduction}

The maturation of open-source systems and in-vehicle computing platforms has shifted research from isolated algorithmic validation toward system-level experimentation involving real vehicles, controlled road environments, and heterogeneous operating conditions. Existing studies have mainly examined functional modules such as perception, localization, planning, and control, and the corresponding experiments are often organized in a modular or localized manner.  However, the operational behavior of an autonomous driving system in a real vehicle cannot be attributed to a single algorithmic module. It is affected by sensor input, vehicle response, onboard computation, and the communication state during a given road scenario. This coupling suggests that single-module or single-scenario experiments are often insufficient for system-level validation, problem tracing, and experimental reuse~\cite{huai2023doppeltest,ding2023safety_critical_scenario_survey}.

Open-source autonomous driving systems provide an inspectable and extensible software foundation for autonomous driving research. Compared with closed systems, open-source systems provide greater transparency in system architecture, module interfaces, and data-flow processes, enabling researchers to examine the implementation logic and operating boundaries of autonomous driving functions from a system perspective. Once an open-source system is deployed on a real-vehicle platform, its research value extends beyond algorithm-function validation to the analysis of system operation, experimental constraints, and multi-factor coordination. A stable reusable research environment with traceable experimental conditions is a prerequisite for realizing the research value of open-source autonomous driving systems.

Nevertheless, the integration of open-source autonomous driving systems with real-vehicle platforms does not itself constitute a reproducible research methodology~\cite{song2024scenario_based_testing_path}. Existing real-vehicle experiments are often limited to a single vehicle or specific scenario, with task continuity across teams not always well maintained. Code, configurations, scripts, and platform-use experience generated during experiments are often difficult to reproduce or extend in the absence of a unified sharing mechanism~\cite{chen2024misconfiguration_ads,chen2025bug_fix_patterns_ads}. The connection between software runtime analysis and hardware-side experimentation also remains insufficient, while real-vehicle validation and external feedback are not always incorporated into a sustained improvement process. These limitations indicate that the research value of an open-source autonomous-driving platform is also shaped by how experiments are organized, how evidence is retained, and how testing results are reviewed after vehicle operation.

In response, this paper examines an Apollo-on-Hongqi EV research environment based on Baidu-donated test vehicles and the Apollo open-source system~\cite{apollo_open_autonomous_driving_platform}. The donated vehicle platform makes it possible to connect real-vehicle experimentation with Apollo-based recording and later testing work in a shared university-industry research setting. Within this setting, experimental materials are retained together with the conditions under which they are produced, so that results from one task can be reviewed in relation to subsequent research activities. Based on this environment, the paper develops a hardware-in-the-loop (HIL) research framework in which multi-vehicle experimentation, code reuse, and software-hardware testing are organized as connected parts of the same review process. Recent studies on scenario construction and industry-oriented autonomous-driving-system testing provide the basis for the testing-support component, which supports evidence-linked review and simulation-based candidate preparation~\cite{tu2025trafficcomposer,song2026rapid_review_ads_testing}. These mechanisms support process tracing, result rechecking, and cross-team reuse in real-vehicle experiments.

The main contributions of this paper are summarized as follows:

\begin{itemize}
    \item The Apollo-on-Hongqi EV environment is characterized as a shared real-vehicle research setting based on Baidu-donated Hongqi EV test vehicles and the Apollo open-source autonomous-driving system. The discussion focuses on how vehicle runs are retained together with their experimental conditions for later review.

    \item A real-vehicle HIL research framework is developed to connect multi-vehicle experimentation, Apollo code reuse, and software--hardware testing within the same review process. LLM-assisted evidence structuring and reinforcement-learning-based scenario generation are incorporated to support traceable record review, anomaly-evidence organization, and candidate-scenario preparation.
    
    \item Preliminary cases from vehicle operation, perception-code adaptation, and hardware-side testing are used to discuss how experimental records can remain reviewable across a shared university-industry research process under retained real-vehicle conditions.

\end{itemize}

The remainder of this paper is organized as follows. Sect.~\ref{sec:related_work} reviews related work. Sect.~\ref{sec:platform_evidence_chain} introduces the composition and operating chain of the Apollo platform deployed on the Hongqi EV. Sect.~\ref{sec:mechanisms} proposes the multidisciplinary HIL research paradigm and its core mechanisms. Sect.~\ref{sec:preliminary_results} analyzes the preliminary experimental results. Sect.~\ref{sec:discussion_conclusion} presents the discussion and conclusion.

\section{Related Work}
\label{sec:related_work}

\subsection{Open-Source Autonomous Driving Platforms and Real-Vehicle Deployment}

As autonomous driving research shifts from single-algorithm evaluation toward system-level experimentation, open-source autonomous driving platforms have gradually become a basis for connecting algorithm development with real-vehicle validation. Zhao et al. reviewed autonomous driving frameworks and simulators, and pointed out that Apollo and Autoware can provide development environments and validation conditions for modular autonomous driving systems, while real hardware access still constrains the expansion of platform applications, and complex scenario construction and co-simulation also affect the completeness of system validation \cite{zhao2024ad_frameworks_simulators}. Aliane further analyzed the research impact of open-source autonomous driving systems, emphasizing their role in cross-institutional collaboration and experimental resource reuse, and discussing their development trends when integrated with artificial intelligence methods and edge-computing conditions \cite{aliane2025oss_ads_survey}. Existing surveys regard open-source platforms as reusable software environments that support system integration and experimental review.

Among representative open-source autonomous driving systems, Apollo and Autoware have received sustained attention. Jung et al. compared their core modules and middleware communication mechanisms, and pointed out that Autoware facilitates research development and functional replacement, whereas Apollo achieves lower latency in large-scale data transmission through Cyber RT and shared-memory mechanisms \cite{jung2025autoware_apollo_compare}. This comparison highlights the value of Apollo as a real-vehicle software foundation. Apollo supports module coordination during vehicle operation through full-stack module organization and real-time communication mechanisms, while its state recording and replay analysis functions provide a basis for post-experiment review. Sensor inputs and onboard computing processes can therefore be examined within the same software chain.

Real-vehicle deployment requires a connection between the open-source software stack and the vehicle execution system, while simulation environments reduce the validation cost before vehicle-side testing. Guo et al. verified the feasibility of connecting an open-source autonomous driving software stack with a drive-by-wire vehicle system \cite{guo2024autoware_dataspeed_dbw}. W\"ursching et al. and Kaljavesi et al. respectively studied interfaces between Autoware and CommonRoad and between Autoware and CARLA, reducing the cost of moving from benchmark testing or simulation environments into complete software-stack testing \cite{wuersching2024cr2aw,kaljavesi2024carla_autoware_bridge}. Open platforms have also been further applied to functional extension and system-operation analysis: Gulzar et al. implemented yielding behavior in Autoware and evaluated it with simulation and real traffic data \cite{gulzar2024roundabouts_unprotected_turns}; Lucchetti et al. integrated resilience mechanisms into Apollo and analyzed the fault-recovery process in a simulation environment \cite{lucchetti2023resilient_apollo}.

Existing literature provides a basis for understanding the system value of open-source autonomous driving platforms. For continuous real-vehicle research, a software change needs to be interpreted together with the condition under which it is executed. A code branch that changes an Apollo module can be reviewed more reliably when its configuration state, runtime record, and replay material are retained with the submitted modification. This relation is important for later reuse, because a result observed in one vehicle run may depend on the activation condition of the code and the recorded Apollo execution state. Taking Apollo-on-Hongqi EV as the object, this paper further examines how the Apollo software chain can support reviewable experimental units and code reuse under real-vehicle conditions.

\subsection{Validation Before Real-Vehicle Operation and Scenario-Based Testing of Autonomous Driving Systems}

Validation under real-vehicle conditions is challenging using only offline data or pure simulation. Offline records can support replay-based analysis of perception, prediction, and planning modules, but scenario coverage is limited by the original data-collection conditions. Simulation environments provide controllability and repeatability; however, differences in vehicle dynamic response and execution delay can still affect the credibility with which test results are transferred to real vehicles. Therefore, validation research before real-vehicle operation has gradually turned to intermediate validation environments such as HIL, vehicle-in-the-loop (ViL), and broader X-in-the-loop (XiL), so that real hardware states and software execution processes can participate in system testing under controlled conditions.

In the construction of testing environments before real-vehicle operation, existing studies mainly focus on the credible relationship between the test environment and real vehicle operation. AD-VILS connects a high-fidelity simulation environment with a real vehicle and evaluates the reliability of the vehicle-in-the-loop platform by comparing virtual test results with real-road test results \cite{oh2024ad_vils}. X-in-the-loop reliability evaluation further incorporates hardware conditions, software execution, and virtual models into a unified testing framework, and analyzes the relationship between simulation testing and real testing from the perspectives of parameter consistency and scenario consistency \cite{oh2024xil_reliability}. Vehicle-in-the-loop research for CCAM environments connects traffic simulation, V2X communication, and a ROS2-based experimental vehicle, and verifies vehicle-in-the-loop testing under connected traffic conditions using adaptive cruise control as an example \cite{coppola2025ccam_vil}. These studies provide a reliability-evaluation basis for validation before real-vehicle operation, while the transferability of results to real vehicles still depends on the specific test platform and scenario conditions.

The effectiveness of validation before real-vehicle operation also depends on whether the test materials can continue to be used in subsequent experiments. DeepScenario constructs an open driving scenario dataset for autonomous driving system testing, enabling executable scenarios and test results to enter subsequent analysis through replay tools \cite{lu2023deepscenario}. The implication of this study is that the research value of test materials depends on stable associations among scenario definitions, operating conditions, and result attributes. For real-vehicle research platforms, a similar logic can be extended to the preservation of associations between vehicle-operation evidence and software-configuration states. Only when experimental results can be traced back to the corresponding scenario and configuration conditions can a single test be transformed into research material that subsequent teams can review and inherit.

In recent years, intelligent methods have begun to be used for scenario generation, test-material organization, and boundary-condition search before real-vehicle operation~\cite{yang2025critical_pedestrian}. ChatScene uses a large language model (LLM) and a knowledge-retrieval mechanism to transform natural-language traffic scenario descriptions into executable Scenic scripts in CARLA, thereby providing a text-to-simulation conversion path for safety-critical scenario generation \cite{zhang2024chatscene}. Research on safety-case generation for Baidu Apollo shows that LLMs can assist in organizing safety-argument materials in an autonomous-driving software stack under domain evidence and human-review constraints \cite{odu2025llm_safety_case_apollo}. Reinforcement-learning-based methods have also been used for boundary-scenario search~\cite{li2022multiagent_signal,yang2026task_scheduling}. A safety-critical scenario generation method based on reinforcement-learning editing searches for high-risk scenarios by adding traffic participants and modifying trajectories, while considering both risk and plausibility in reward design \cite{liu2023rl_scenario_editing}. The DenseRL-based adaptive testing environment generation method improves the evaluation efficiency of connected and automated vehicles by learning critical scenarios in which the surrogate model and the system under test differ substantially \cite{yang2025denserl_testing}. These studies show that LLMs can support the structuring of scenario descriptions and evidence-related materials, while reinforcement-learning-based methods can explore candidate boundary conditions before vehicle-side verification. The generated scenarios and boundary candidates still need to be reviewed against real-vehicle records, domain knowledge, and safety constraints before being used in subsequent experiments.

Existing studies have advanced pre-deployment validation before real-vehicle operation by improving test-environment credibility and scenario preparation. For a continuously used real-vehicle research platform, validation results still need to be retained with the vehicle condition and software state that produced them. This requirement is closely related to later code review, software-hardware testing, and cross-team examination of experimental results. Taking the Apollo-on-Hongqi EV platform as the object, this paper focuses on how real-vehicle records, code-submission materials, and testing feedback can be organized into reviewable research evidence under multi-vehicle and multidisciplinary experimental conditions.

\section{Apollo-on-Hongqi EV Platform and Operational Evidence Chain}
\label{sec:platform_evidence_chain}

\begin{figure}[t]
\centering
\includegraphics[width=0.96\textwidth,trim=8 8 8 8,clip]{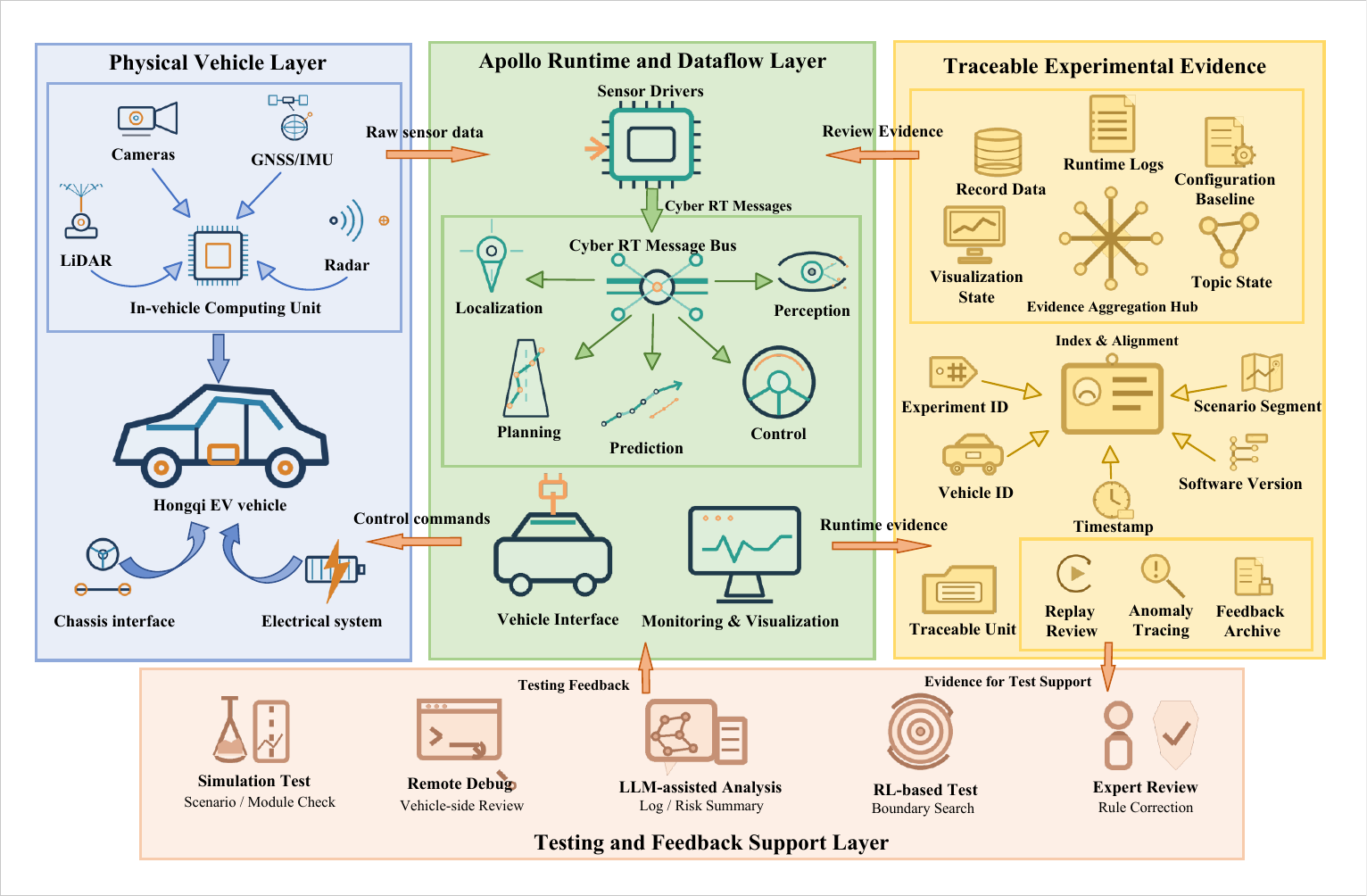}
\caption{Reviewable operational data chain of the Apollo-on-Hongqi EV platform.}
\label{fig:apollo_hongqi_overall_architecture}
\end{figure}

In the Hongqi EV real-vehicle environment, the Baidu Apollo open-source autonomous driving system is analyzed through the records generated during vehicle operation. As shown in Fig.~\ref{fig:apollo_hongqi_overall_architecture}, the Hongqi EV provides the operating context, while Apollo records the runtime process associated with the observed system response. The analysis focuses on how vehicle operation, Apollo runtime records, and post-run review materials are linked after each run.

Fig.~\ref{fig:apollo_hongqi_overall_architecture} describes how a real-vehicle run is retained for later review. The Hongqi EV provides the operating context in which Apollo produces its runtime response, and the vehicle-side condition is kept together with the observed result. During the run, Apollo records the software process associated with this context, so that the response can later be replayed and checked against the original operating condition. After the run, the retained record is used when examining code changes, hardware-side tests, or scenario-related observations. In this way, the figure shows how vehicle operation, Apollo recording, and later testing review remain connected within the same experimental evidence chain.

During a vehicle run, Apollo receives real-vehicle inputs and generates corresponding runtime responses under the current operating condition. The recorded material preserves the context in which the observed response occurred. Later analysis can therefore return to the execution context of the original run when examining an experimental result.

The relation between an observed phenomenon and its operating context is important for real-vehicle research. A change in the test environment or Apollo configuration may affect runtime behavior. The communication state during the run may also influence the continuity of the retained material. For this reason, each experimental run is associated with its corresponding Apollo record. This association provides the basis for tracing a result back to the condition under which it was generated.

In Apollo, Cyber RT provides the communication framework for module interaction during system operation \cite{apollo_cyber_rt_framework}. In this paper, the recorded communication process is used to relate Apollo runtime observation to the corresponding vehicle run. Its value lies in preserving the contextual basis required for later interpretation.

For experimental review, a run-level index is used to identify the Apollo record that belongs to a given test period. The index links the retained record to the vehicle run in which it was produced and preserves the information required to locate the corresponding software state and scenario segment. With this relation, the retained material can be reviewed under the execution context of the original run.

Apollo's record and replay capability supports this review process \cite{apollo_cyber_developer_tools}. Replay enables researchers to examine the recorded communication process after the vehicle test has ended. In this study, replay is used to check whether an observed response is consistent with the retained runtime material and to judge whether the result can serve as a reference for later experimental comparison.

The Apollo-on-Hongqi EV records keep each experimental result connected with the vehicle condition, software state, and scenario segment in which it was generated. This connection gives post-run review a concrete reference and helps prepare subsequent HIL experiments. In this way, a completed vehicle run can be replayed and compared under its original operating conditions.

In summary, the Apollo-on-Hongqi EV platform provides an evidence basis for the experimental framework. Apollo runtime records can be interpreted in relation to the real-vehicle condition and configuration state of the original run. This evidence chain supports the collaborative research framework discussed in the next section.

\section{Proposed Multi-Disciplinary Hardware-in-the-Loop Research Paradigm}
\label{sec:mechanisms}

\begin{figure}[t]
\centering
\includegraphics[width=0.96\textwidth,trim=1 1 1 1,clip]{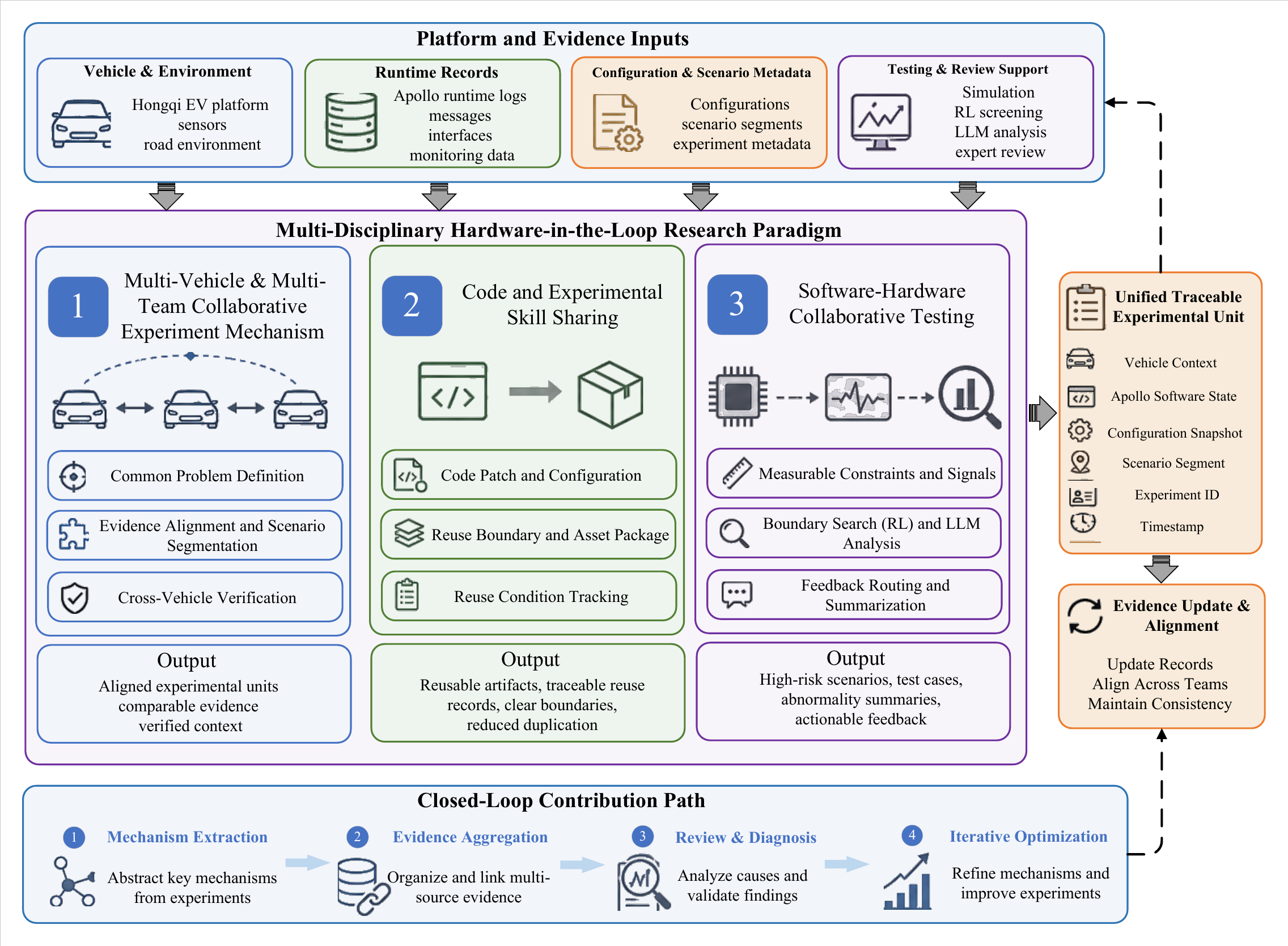}
\caption{Evidence-driven multidisciplinary HIL research paradigm for Apollo-on-Hongqi EV.}
\label{fig:paradigm}
\end{figure}

Based on the Hongqi EV environment and the Apollo operating chain described above, the proposed paradigm organizes each retained vehicle run as a traceable experimental unit \(\mathcal{U}_i=\langle v_i,s_i^{\mathrm{sw}},c_i^{\mathrm{cfg}},x_i^{\mathrm{scn}},d_i^{\mathrm{rec}},t_i,e_i^{\mathrm{rev}}\rangle\). In this unit, \(v_i\) identifies the vehicle condition, \(s_i^{\mathrm{sw}}\) the Apollo software state, \(c_i^{\mathrm{cfg}}\) the configuration snapshot, \(x_i^{\mathrm{scn}}\) the scenario segment, \(d_i^{\mathrm{rec}}\) the runtime record, \(t_i\) the time index, and \(e_i^{\mathrm{rev}}\) the retained review information. As shown in Fig.~\ref{fig:paradigm}, multi-vehicle collaboration aligns such units across vehicle and team conditions, code reuse preserves the activation conditions and reuse boundaries associated with each unit, and software--hardware collaborative testing attaches model-assisted review records and candidate scenarios to the corresponding unit.

\subsection{Mechanism 1: Multi-Vehicle and Multi-Team Collaborative Experiment Mechanism}

The multi-vehicle and multi-team collaborative experiment mechanism is designed for shared real-vehicle research with Hongqi EVs equipped with Apollo. In this environment, control, electrical, cybersecurity, and perception-related observations may be made on the same vehicle run or on comparable runs across vehicles. A real-vehicle phenomenon is usually affected by vehicle motion, sensor input, Apollo configuration, onboard computation, and Cyber RT communication. If each team interprets the phenomenon only from its own record, a coupled software-hardware issue may be simplified into a single-module problem. For this reason, collaborative experiments should be organized around the same experimental problem, scenario segment, and Apollo runtime record. Within this organization, the experiment log keeps the task under examination connected with the vehicle condition and the retained control process, allowing related runs to be reviewed under their recorded conditions. Later code reuse and software-hardware testing can then return to this run-level context when an observed result needs to be interpreted.

In a low-speed campus road test, the common experimental problem can be defined around the operating stability of Apollo under real sensor input and chassis response. The control direction can examine planned trajectories, control commands, and vehicle responses. The electrical direction can check sensor access, device status, and onboard interface conditions. The cybersecurity direction can review Cyber RT topic continuity, communication state, and abnormal logs~\cite{miao2025adaptive_sensor_attack}. These observations are meaningful when they are linked to the same vehicle operation process and the same recorded data segment. In this way, different disciplinary observations can be compared on a common experimental basis.

During a real-vehicle experiment, analysis should be associated with specific test segments or abnormal events. For trajectory deviation, the planned trajectory, control output, localization state, chassis response, and communication delay should be checked together~\cite{xiao2025offroute,wang2025mhtraj}. For perception-target loss, the analysis should return to sensor input, perception output, topic state, and record completeness. For module timeout, computing status, runtime logs, communication state, and startup configuration should be reviewed together. Associating these materials with the same event reduces the risk of judging a coupled system phenomenon from a single module output.

Configuration records are necessary for maintaining continuity across different runs. Sensor parameters, vehicle parameters, startup files, topic settings, and record channels can affect Apollo input, module behavior, and log output. If these changes are separated from experimental results, later teams may attribute configuration-induced phenomena to algorithm behavior or vehicle state. Each experiment should therefore retain a configuration baseline and record task-related changes, affected modules, and applicable conditions. This allows subsequent analysis to distinguish vehicle differences, configuration changes, and software behavior.

When the same experimental problem is extended to multiple Hongqi EVs, the purpose is to compare whether the observed phenomenon changes under different real-vehicle conditions. A baseline observation can be obtained on one vehicle, and a comparable run can then be conducted on another vehicle under similar scenario and configuration conditions. If similar phenomena appear across vehicles, the observation can be compared across vehicles under explicit conditions. If the results differ, the analysis should return to vehicle-side status, sensor configuration, software version, and runtime records. In this way, the condition sensitivity of the observed phenomenon can be examined across different real-vehicle conditions.

Multidisciplinary observations and cross-vehicle comparison records, together with expert comments, are attached to the corresponding \(\mathcal{U}_i\), allowing later teams to replay and compare the run under its retained operating conditions. When the same experimental problem is revisited, this association supports configuration rechecking and abnormal-phenomenon interpretation before follow-up correction plans are made.

In summary, this mechanism organizes each real-vehicle experiment as a traceable unit defined by a shared experimental problem and the conditions recorded during the run. The unit keeps the observed vehicle response connected with the corresponding Apollo runtime record, so that later review can examine the phenomenon under its original operating condition. This structure provides the evidence basis for subsequent code reuse and software-hardware testing.

\subsection{Mechanism 2: Code and Experimental Skill Sharing Mechanism}

The reproducibility of Apollo real-vehicle experiments largely depends on whether a code change remains tied to the condition in which it was tested. On the Hongqi EV platform, the repository keeps a submitted branch linked to its Apollo execution record and the experimental condition under which the branch is activated. Later researchers can use this relation to examine the startup of the related module and the publication of expected Cyber RT topics in the original vehicle-side context. In this way, each code modification can be traced back to its test condition during later review and reuse.

In the proposed framework, code sharing is organized around the correspondence among software changes, vehicle operating conditions, and validation evidence. The code repository preserves identifiable software modifications and necessary non-sensitive configuration materials, while the related experimental context and review evidence are supplemented through internal knowledge materials. These two parts are connected through experimental asset packages, so that code outcomes can be preserved together with their applicable conditions, runtime records, and review information. Table~\ref{tab:experiment_asset_package} shows how the use of a submitted branch is described through its activation condition, execution record, and reuse boundary. Through this organization, subsequent teams can understand the generation conditions and validation basis of the code during reuse, thereby reducing semantic loss and reproduction risks in cross-vehicle and cross-task migration.

\begin{table}[!htbp]
\caption{Evidence Structure of Experimental Asset Packages in the Code-Sharing Mechanism}
\label{tab:experiment_asset_package}
\centering
\footnotesize
\setlength{\tabcolsep}{5pt}
\renewcommand{\arraystretch}{1.18}
\begin{tabularx}{\textwidth}{@{}m{0.18\textwidth}m{0.31\textwidth}X@{}}
\toprule
Evidence Layer & Main Representation & Supporting Role \\
\midrule
Software state
& Branch identifier, Apollo module change, and auxiliary script
& Identifies the system position of code changes and supports traceability. \\

Execution condition
& Vehicle and sensor settings, module parameters, and software version
& Specifies the constraints under which experimental results are generated. \\

Runtime observation
& Record index, log summary, topic state, and visualization output
& Converts code execution into inspectable runtime evidence for review. \\

Experience interpretation
& Operating path, debugging note, risk cue, and rollback basis
& Transforms tacit experimental experience into transferable interpretation. \\

Sharing boundary
& Public materials, internal materials, and sensitive-data rules
& Delimits open collaboration materials and controlled internal materials. \\
\bottomrule
\end{tabularx}
\end{table}

When a new reuse task is submitted, LLM-assisted retrieval relates its description to earlier experimental asset packages with comparable Apollo execution evidence. The retrieved material keeps the branch source and the operating condition needed for reuse checking. Its verification record is used to recover the original execution context, so that researchers can judge whether the earlier code outcome matches the target vehicle condition. The next subsection formulates this reuse-oriented application as the experimental-asset retrieval task \(\mathcal{L}_{\mathrm E}\).

Selected code outcomes are then rechecked in a controlled vehicle-side debugging environment. The review confirms whether the material enters the expected Apollo runtime path and whether the required data links are preserved during execution. Abnormal logs and retained verification records are examined together before the material is considered for migration to another vehicle or experimental task.

The release scope depends on the reproduction value of the material and the constraint level of the vehicle test. Code changes with clear operating conditions and sufficient verification evidence can enter code sharing or open-source collaboration. Materials bound to vehicle parameters, real-environment records, or sensitive configurations remain under controlled handling. LLM-assisted review is used to flag incomplete reproduction descriptions, while researchers decide the final release scope and pull-request submission.

Code sharing is treated as a condition-dependent part of Apollo real-vehicle experimentation. A submitted branch is reviewed together with the configuration switch that activates it and the runtime record generated during execution, so that later researchers can judge whether the same modification can be used on a target Hongqi EV. When the branch is adapted to another task, the related startup condition and verification result are used to confirm the expected module behavior, data-link status, and record output before vehicle-side use. Corrections obtained during this process update the corresponding experimental asset package and clarify its applicable scope. The sharing mechanism supports controlled reuse of Apollo code outcomes under explicit experimental conditions.

\subsection{Mechanism 3: Software-Hardware Collaborative Testing Mechanism}

Within the software-hardware collaborative testing mechanism, LLM-assisted processing organizes experimental evidence into traceable task records, while reinforcement learning searches for candidate simulation scenarios under task-specific constraints. Both processes retain their association with the original vehicle record and support researcher review during test preparation.

This mechanism starts from the correspondence between real-world constraints and Apollo runtime evidence. Vehicle-side operation provides physical constraints, while Apollo logs, record replay, and DreamView visualization provide observable software evidence. Expert review is introduced to constrain uncertain interpretations produced during automated or data-driven analysis.

\begin{figure}[!t]
\centering
\begin{minipage}{\textwidth}
    \centering
    \includegraphics[width=\textwidth,trim=4 4 4 4,clip]{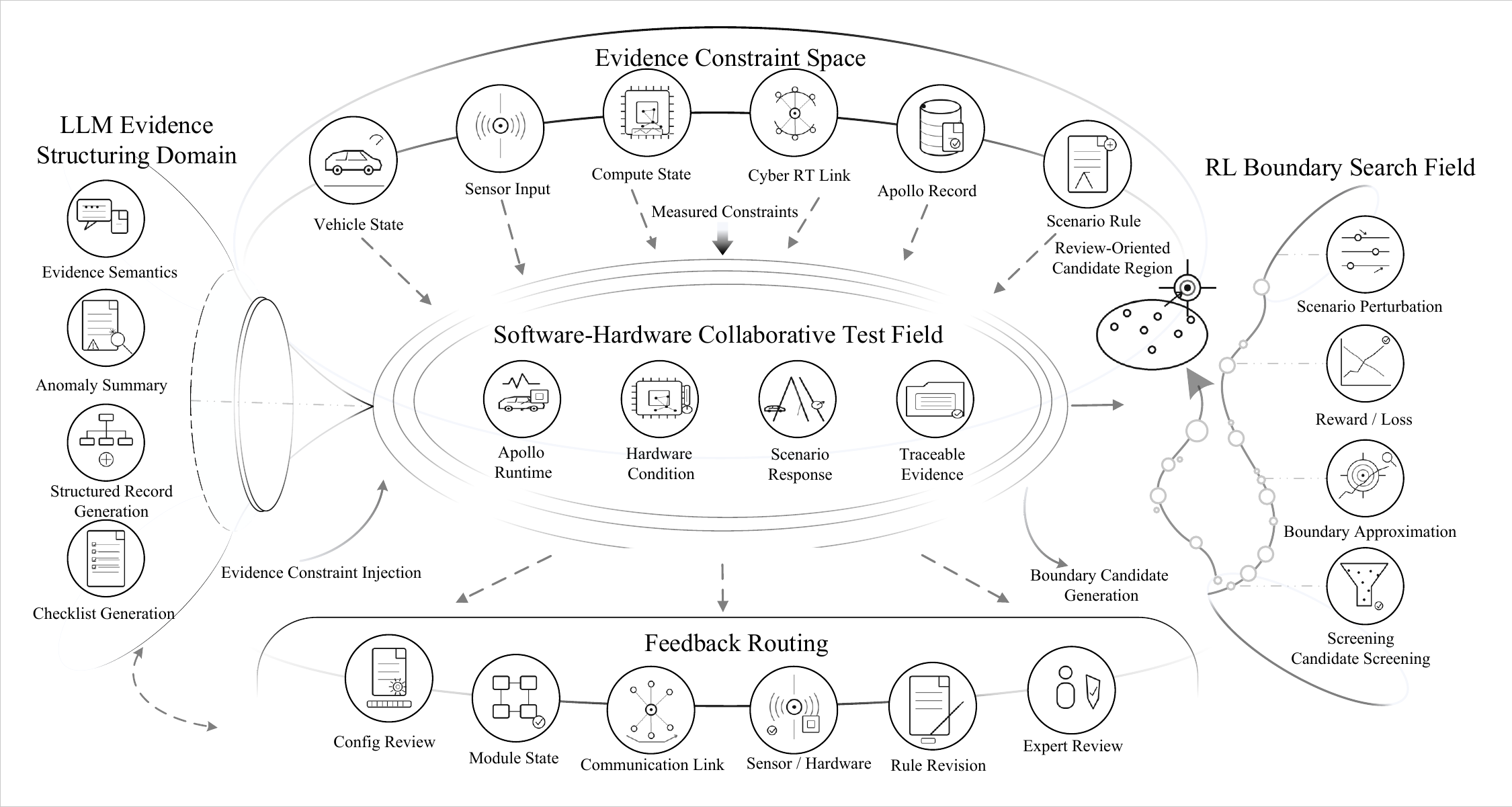}
    \par\vspace{1pt}
    \small (a) Overall software--hardware collaborative testing mechanism.
\end{minipage}

\vspace{4pt}

\begin{minipage}[t]{0.49\textwidth}
    \centering
    \includegraphics[width=\linewidth,trim=4 4 4 4,clip]{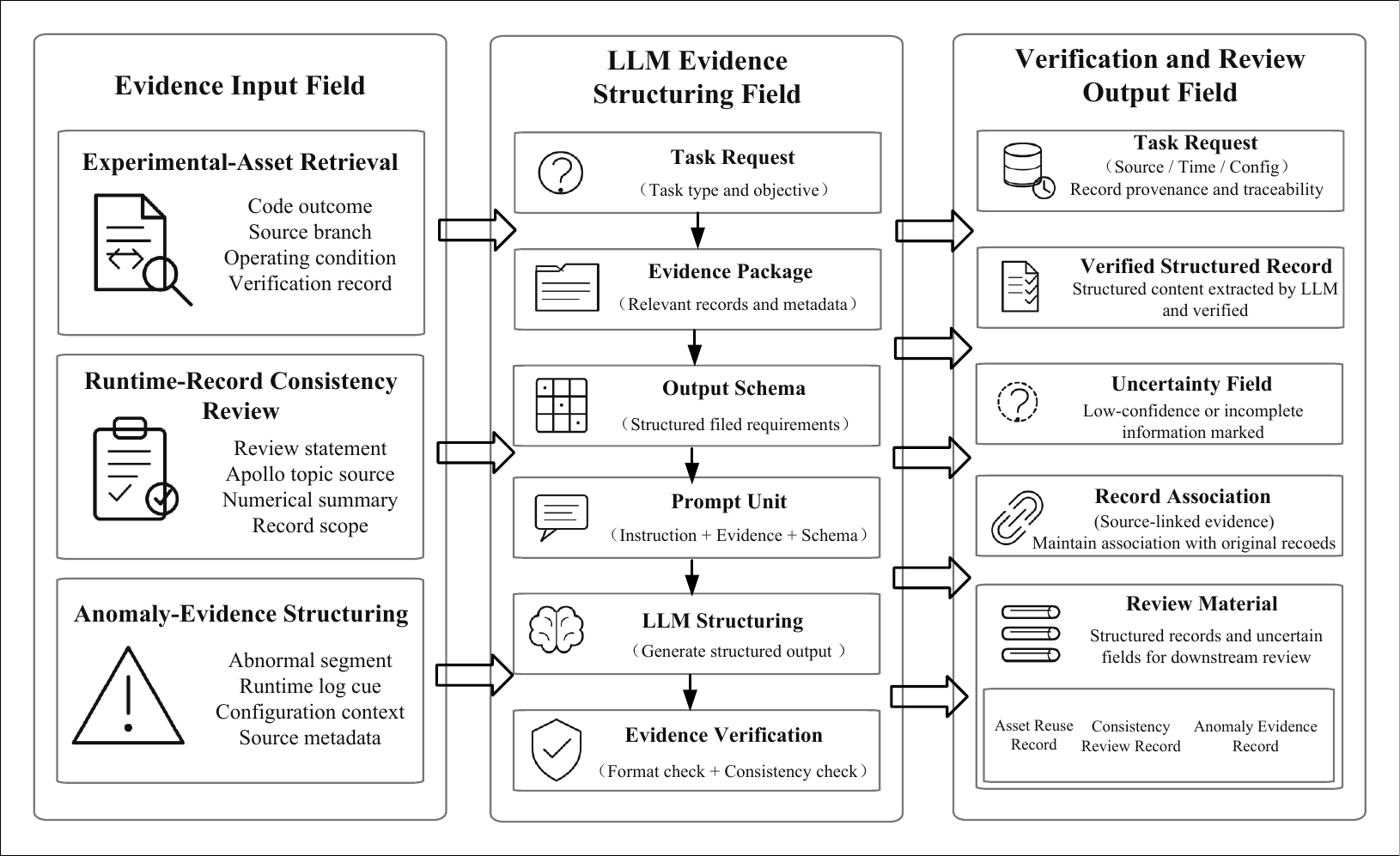}
    \par\vspace{1pt}
    \small (b) Task scenarios for LLM-assisted evidence processing.
\end{minipage}\hfill
\begin{minipage}[t]{0.49\textwidth}
    \centering
    \includegraphics[width=\linewidth,trim=4 4 4 4,clip]{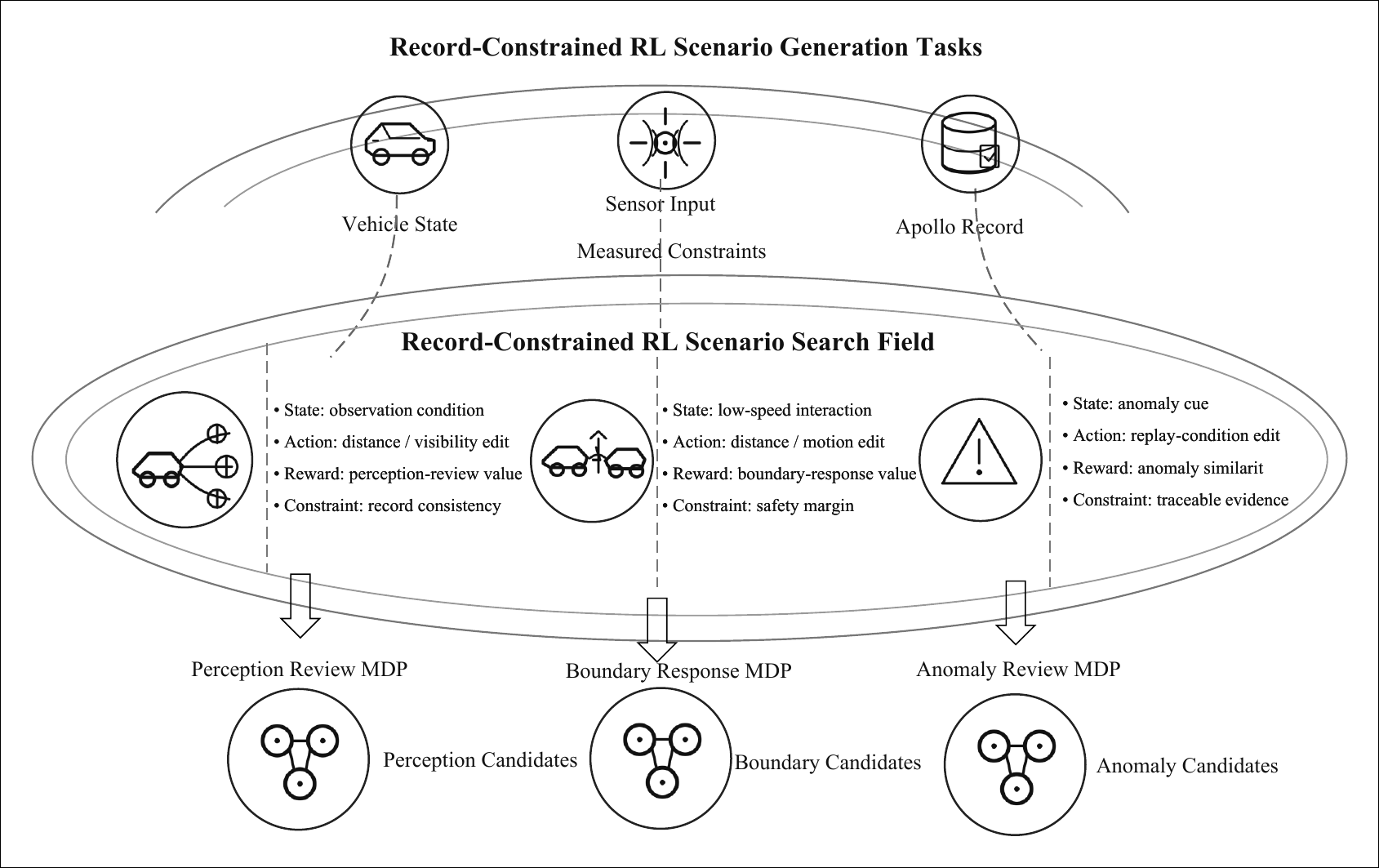}
    \par\vspace{1pt}
    \small (c) Task scenarios for reinforcement-learning-based scenario generation
\end{minipage}
\caption{Evidence-constrained software--hardware collaborative testing mechanism: (a) overall mechanism; (b) task scenarios for LLM-assisted evidence processing; (c) task scenarios for reinforcement-learning-based scenario generation.}
\label{fig:sh_intelligent_loop}
\end{figure}

Fig.~\ref{fig:sh_intelligent_loop}(a) presents the overall evidence-constrained testing mechanism. Vehicle-side operating evidence delimits the admissible test conditions, while LLM-assisted rule formation and reinforcement-learning-based boundary search provide complementary inputs to the software--hardware collaborative test field. Reviewed outputs are returned through the feedback route so that test constraints and subsequent preparation can be revised while remaining linked to the original vehicle record.

\subsubsection{LLM-Assisted Evidence Structuring for Review}
\label{sec:llm_evidence_review}
Fig.~\ref{fig:sh_intelligent_loop}(b) abstracts LLM-assisted evidence structuring. Vehicle-side evidence is organized according to three review needs. Prior Apollo code assets require reuse checking, generated statements are checked against retained runtime records, and abnormal segments are anchored before replay or reconstruction. These needs define \(\mathcal{L}_{\mathrm E}\), \(\mathcal{L}_{\mathrm R}\), and \(\mathcal{L}_{\mathrm A}\). For task \(j\in\{\mathrm E,\mathrm R,\mathrm A\}\), the processing relation is \(\mathcal{L}_j:(q_j,\mathcal{E}_j,\Pi_j)\mapsto(o_j,u_j)\), where \(q_j\) is the task request, \(\mathcal{E}_j\) is the retained evidence, and \(\Pi_j\) specifies the output-field set \(\mathcal{F}_j\). The verified result partitions these fields according to \(\mathcal{F}_j=o_j\cup u_j\) and \(o_j\cap u_j=\varnothing\). Fields traceable to \(\mathcal{E}_j\) enter \(o_j\), while missing, conflicting, or unsupported fields enter \(u_j\) for researcher inspection.

\paragraph{Experimental-Asset Retrieval Task \(\mathcal{L}_{\mathrm E}\).}
The experimental-asset retrieval task retrieves earlier Apollo code outcomes that may support a new vehicle or experimental task. The request \(q_{\mathrm E}\) specifies the target vehicle condition, reuse purpose, and expected Apollo runtime behavior, while \(\mathcal{D}_{\mathrm E}\) denotes the repository of evidence-linked asset packages. The retrieved evidence set is written as \(\mathcal{E}_{\mathrm E}(q_{\mathrm E})=\operatorname{TopK}_{K_{\mathrm E}}(\mathcal{D}_{\mathrm E};\operatorname{Score}_{\mathrm E}(q_{\mathrm E},\cdot))=\{e_i\}_{i=1}^{K_{\mathrm E}}\), where \(\operatorname{Score}_{\mathrm E}\) measures request--asset compatibility under the retained operating and execution conditions, and \(K_{\mathrm E}\) is the number of candidates retained for review. Each retrieved \(e_i\) preserves its source branch, operating condition, and verification record.

For each retrieved candidate \(e_i\), the model input is constructed as \(x_i^{\mathrm E}=\operatorname{Prompt}_{\mathrm E}(q_{\mathrm E},e_i,\Pi_{\mathrm E})\), where \(\Pi_{\mathrm E}\) specifies the reuse-record fields and traceability rules. The language model produces a candidate reuse record and an unresolved field set, written as \((\hat{o}_i^{\mathrm E},\hat{u}_i^{\mathrm E})=\operatorname{LLM}_{\psi}(x_i^{\mathrm E})\). The record \(\hat{o}_i^{\mathrm E}\) preserves the source association, applicable condition, expected Apollo runtime path, and verification basis, while \(\hat{u}_i^{\mathrm E}\) keeps fields that are missing, conflicting, or unsupported by \(e_i\).

The generated record is verified against the original candidate evidence by \((o_i^{\mathrm E},u_i^{\mathrm E})=\operatorname{Verify}_{\mathrm E}(\hat{o}_i^{\mathrm E},\hat{u}_i^{\mathrm E};e_i)\). The verification operation retains a statement in \(o_i^{\mathrm E}\) only when it can be traced to \(e_i\), and unresolved fields are kept in \(u_i^{\mathrm E}\) for researcher inspection. The verified result supports the subsequent reuse decision: candidates with sufficient evidence can enter controlled vehicle-side verification, candidates requiring modification are returned for adaptation, and evidence-insufficient cases remain under controlled storage.

\paragraph{Runtime-Record Consistency Review Task \(\mathcal{L}_{\mathrm R}\).}
The runtime-record consistency review task examines whether a generated review statement is supported by retained vehicle-side evidence. The request \(q_{\mathrm R}\) contains the statement to be checked and defines the record scope against which it is examined. The evidence set \(\mathcal{E}_{\mathrm R}\) contains the time-bounded Apollo record segment and its retained numerical summaries, with source metadata and operating context attached to the same scope. The accepted review record is therefore restricted to content that can be traced to the specified evidence.

The supplied statement is decomposed into independently verifiable units, written as \(\mathcal{Z}_{\mathrm R}=\operatorname{Split}_{\mathrm R}(q_{\mathrm R},\Pi_{\mathrm R})\). Each \(z_m\in\mathcal{Z}_{\mathrm R}\) represents an assertion requiring support from the retained record. Its model input is constructed as \(x_m^{\mathrm R}=\operatorname{Prompt}_{\mathrm R}(z_m,\mathcal{E}_{\mathrm R},\Pi_{\mathrm R})\), where \(\Pi_{\mathrm R}\) specifies the evidence-checking schema and requires the generated result to preserve the available source reference and numerical basis.

For each unit, the language model produces a candidate review statement and a corresponding unresolved field set, expressed as \((\hat{o}_m^{\mathrm R},\hat{u}_m^{\mathrm R})=\operatorname{LLM}_{\psi}(x_m^{\mathrm R})\). The candidate statement \(\hat{o}_m^{\mathrm R}\) contains the record-grounded content proposed by the model, and \(\hat{u}_m^{\mathrm R}\) keeps content that cannot be supported within the supplied evidence scope. The generated fields then enter the evidence-verification operation \((o_m^{\mathrm R},u_m^{\mathrm R})=\operatorname{Verify}_{\mathrm R}(\hat{o}_m^{\mathrm R},\hat{u}_m^{\mathrm R};\mathcal{E}_{\mathrm R})\). This operation matches accepted statements to their source evidence and leaves unresolved content in \(u_m^{\mathrm R}\) for researcher inspection.

The verified outputs are collected as \(o_{\mathrm R}=\{o_m^{\mathrm R}\}_{m=1}^{|\mathcal{Z}_{\mathrm R}|}\) and \(u_{\mathrm R}=\{u_m^{\mathrm R}\}_{m=1}^{|\mathcal{Z}_{\mathrm R}|}\). Researchers inspect the unresolved fields in \(u_{\mathrm R}\) before the accepted review record \(o_{\mathrm R}\) is used in subsequent analysis.

\paragraph{Anomaly-Evidence Structuring Task \(\mathcal{L}_{\mathrm A}\).}
The anomaly-evidence structuring task converts an observed abnormal segment into a traceable evidence record for later review. The request \(q_{\mathrm A}\) identifies the anomaly source and defines the record range permitted for examination. The evidence set \(\mathcal{E}_{\mathrm A}\) contains the timestamped runtime material associated with the abnormal segment, while configuration context and source metadata are kept with the retained evidence.

For the \(i\)-th abnormal segment, an anchored evidence package is constructed as \(b_i^{\mathrm A}=\operatorname{Anchor}_{\mathrm A}(q_{\mathrm A},\mathcal{E}_{\mathrm A},w_i)\), where \(w_i\) defines the permitted temporal window. The anchoring operation selects records within this window and preserves their association with the observed anomaly. The model input is then formed as \(x_i^{\mathrm A}=\operatorname{Prompt}_{\mathrm A}(b_i^{\mathrm A},\Pi_{\mathrm A})\), where \(\Pi_{\mathrm A}\) specifies the structure required for an anomaly-evidence record.

Given \(x_i^{\mathrm A}\), the language model produces a candidate record and an unresolved field set, written as \((\hat{o}_i^{\mathrm A},\hat{u}_i^{\mathrm A})=\operatorname{LLM}_{\psi}(x_i^{\mathrm A})\). The candidate record \(\hat{o}_i^{\mathrm A}\) keeps the reconstruction context that can be linked to the anchored evidence, and \(\hat{u}_i^{\mathrm A}\) records content with insufficient support. The generated result is checked against the original evidence package by \((o_i^{\mathrm A},u_i^{\mathrm A})=\operatorname{Verify}_{\mathrm A}(\hat{o}_i^{\mathrm A},\hat{u}_i^{\mathrm A};b_i^{\mathrm A})\). Content traceable to \(b_i^{\mathrm A}\) is retained in \(o_i^{\mathrm A}\), while unresolved content remains in \(u_i^{\mathrm A}\).

The verified outputs are collected as \(o_{\mathrm A}=\{o_i^{\mathrm A}\}_{i=1}^{N_{\mathrm{seg}}}\) and \(u_{\mathrm A}=\{u_i^{\mathrm A}\}_{i=1}^{N_{\mathrm{seg}}}\). The accepted set \(o_{\mathrm A}\) supports subsequent anomaly review and the preparation of reconstruction requests. Researchers inspect \(u_{\mathrm A}\) to determine whether additional evidence is required before the abnormal segment enters later testing.

After the three LLM-assisted tasks have been defined, the next step is to clarify how their inputs are processed into evidence-linked outputs. The experimental-asset task starts from candidate code evidence, the consistency-review task starts from a statement that must be checked within a bounded record scope, and the anomaly task starts from an anchored abnormal segment. Although their evidence units differ, each task keeps the generated record traceable to the retained vehicle-side evidence. Table~\ref{tab:llm_evidence_processing_mapping} presents the corresponding processing flow from request reception to result collection.

\begin{table}[!htbp]
\centering
\caption{Task-specific processing flow for LLM-assisted evidence processing.}
\label{tab:llm_evidence_processing_mapping}
\scriptsize
\setlength{\tabcolsep}{3.2pt}
\renewcommand{\arraystretch}{1.18}

\begin{tabular}{@{}
>{\centering\arraybackslash}m{0.12\textwidth}
>{\centering\arraybackslash}m{0.21\textwidth}
>{\centering\arraybackslash}m{0.22\textwidth}
>{\centering\arraybackslash}m{0.22\textwidth}
>{\centering\arraybackslash}m{0.15\textwidth}
@{}}
\hline

\textbf{Process Node} &
\textbf{Experimental-Asset Retrieval \(\mathcal{L}_{\mathrm E}\)} &
\textbf{Runtime-Record Consistency Review \(\mathcal{L}_{\mathrm R}\)} &
\textbf{Anomaly-Evidence Structuring \(\mathcal{L}_{\mathrm A}\)} &
\textbf{Stage Output} \\
\hline

Request reception &
Code-outcome reuse request; target vehicle condition &
Review statement; record-checking scope &
Anomaly-review request; evidence scope &
\(q_j\) \\
\hline

Evidence scope &
Asset repository \(\mathcal{D}_{\mathrm E}\); execution evidence &
Time-bounded Apollo record; numerical summary &
Anomaly-neighborhood record; runtime context &
\(\mathcal{E}_j\) \\
\hline

Source association &
Source branch; operating condition; verification record &
Topic source; time range; operating condition &
Anomaly time; configuration state; source record &
Source-linked evidence \\
\hline

Task unit &
Retrieved asset \(e_i\) &
Verifiable statement unit \(z_m\) &
Anchored evidence package \(b_i^{\mathrm A}\) &
Task input unit \\
\hline

Output constraint &
Reuse-record fields; traceability rules &
Evidence-checking fields; numerical basis &
Anomaly-record fields; evidence-index rules &
\(\Pi_j\) \\
\hline

Prompt construction &
\(q_{\mathrm E};\,e_i;\,\Pi_{\mathrm E}\) &
\(z_m;\,\mathcal{E}_{\mathrm R};\,\Pi_{\mathrm R}\) &
\(b_i^{\mathrm A};\,\Pi_{\mathrm A}\) &
\(x_\ell^j\) \\
\hline

LLM generation &
Candidate reuse record; fields for checking &
Candidate evidence-linked statement; unresolved fields &
Candidate anomaly record; fields for completion &
\((\hat{o}_\ell^j,\hat{u}_\ell^j)\) \\
\hline

Evidence verification &
\(\operatorname{Verify}_{\mathrm E}(\cdot\,;e_i)\) &
\(\operatorname{Verify}_{\mathrm R}(\cdot\,;\mathcal{E}_{\mathrm R})\) &
\(\operatorname{Verify}_{\mathrm A}(\cdot\,;b_i^{\mathrm A})\) &
\((o_\ell^j,u_\ell^j)\) \\
\hline

Unresolved fields &
Missing evidence; controlled-handling requirement &
Numerical conflict; missing source; out-of-scope content &
Missing log; incomplete configuration; unclear source &
\(u_\ell^j\) \\
\hline

Structured record &
Asset source; applicable condition; verification basis &
Evidence-supported statement; numerical source; runtime context &
Anomaly cue; reconstruction context; evidence index &
\(o_\ell^j\) \\
\hline

Result collection &
\(\{o_i^{\mathrm E}\};\,\{u_i^{\mathrm E}\}\) &
\(\{o_m^{\mathrm R}\};\,\{u_m^{\mathrm R}\}\) &
\(\{o_i^{\mathrm A}\};\,\{u_i^{\mathrm A}\}\) &
\((o_j,u_j)\) \\
\hline

\end{tabular}

\vspace{1mm}

\parbox{0.92\textwidth}{
\centering
\scriptsize
Here, \(j\in\{\mathrm E,\mathrm R,\mathrm A\}\) identifies the processing
task; \(\ell\) denotes its local task-unit index: \(i\) for
\(\mathcal{L}_{\mathrm E}\) and \(\mathcal{L}_{\mathrm A}\);
\(m\) for \(\mathcal{L}_{\mathrm R}\).\par
}

\end{table}

\subsubsection{Reinforcement-Learning-Based Scenario Generation}
\label{sec:rl_scenario_generation}
As shown in Fig.~\ref{fig:sh_intelligent_loop}(c), reinforcement-learning-based scenario search is connected to retained evidence through an evidence--scenario coupling layer. Record anchors and injected constraints determine the admissible context; scenario edits are evaluated through simulation transitions and task feedback, after which screened candidates and their generation trajectories are returned as review material. Within the proposed framework, reinforcement learning supports the construction
of candidate scenarios under constraints supplied by vehicle records. A change
in sensing conditions calls for a perception-review formulation. Evidence
obtained near a low-speed interaction limit directs the search toward system
response. An abnormal episode supplies the reference required for controlled
reconstruction. These review purposes define the formulations introduced below.

For \(k\in\{\mathrm P,\mathrm B,\mathrm A\}\), the corresponding search problem is represented by the MDP
\begin{equation}
\mathcal{M}_{k}
=
\left\langle
\mathcal{S}_{k},
\mathcal{A}_{k},
\mathcal{P}_{k},
\mathcal{R}_{k},
\gamma
\right\rangle .
\end{equation}
Here, \(\mathcal{S}_{k}\) is the scenario state space, \(\mathcal{A}_{k}\) is the scenario-editing action set, \(\mathcal{P}_{k}\) is the simulation transition function, \(\mathcal{R}_{k}\) is the task reward function, and \(\gamma\) is the discount factor. Record and test constraints are defined separately through the feasible trajectory set \(\Omega_k=\{\tau_k=(s_0^k,a_0^k,\ldots,s_T^k)\mid h_{k,r}(\tau_k)\leq0,\ r=1,\ldots,m_k\}\), where \(h_{k,r}\) is the \(r\)-th task-specific constraint and \(m_k\) is the number of constraints defined for task \(k\). The retained candidate set is written as \(\mathcal{C}_k=\{c(\tau_i^k)\mid \tau_i^k\in\Omega_k,\ i=1,\ldots,N_k\}\), where \(c(\tau_i^k)\) maps a feasible trajectory to the corresponding scenario candidate.

Each formulation contains a discrete set of scenario edits, making a deep Q-network (DQN) suitable for action-value learning~\cite{li2023robust_rl}. The common decision and update relations are
\begin{equation}
\begin{aligned}
a_t^k
&\sim
\pi_{\epsilon,k}
\bigl(\cdot\mid s_t^k;Q_{\theta_k}\bigr),\\
\delta_t^k
&=
g_k(s_{t+1}^k),\\
y_t^k
&=
r_t^k+
\gamma(1-\delta_t^k)
\max_{a'\in\mathcal{A}_k}
Q_{\bar{\theta}_k}(s_{t+1}^k,a'),\\
\mathcal{L}_k(\theta_k)
&=
\mathbb{E}_{\mathcal{B}_k}
\left[
\left(
y_t^k-
Q_{\theta_k}(s_t^k,a_t^k)
\right)^2
\right].
\end{aligned}
\end{equation}
Here, \(\pi_{\epsilon,k}\) is the \(\epsilon\)-greedy policy, \(\theta_k\) and \(\bar{\theta}_k\) are the parameters of the online and target networks, \(\mathcal{B}_k\) is the replay buffer, and \(\delta_t^k\in\{0,1\}\) is the termination flag. The target parameters are periodically updated from the online network. Across the three tasks, positive reward components are normalized to \([0,1]\), task-specific weights are nonnegative, and penalty terms such as \(R_s\) and \(R_{\xi}\) are nonnegative. Only candidates generated from feasible trajectories are retained for researcher review.

\paragraph{Perception-Review Scenario Generation \(\mathcal{M}_{\mathrm P}\).}

Perception-review scenario generation addresses perception differences caused by changes in observation conditions during real-vehicle operation. During Apollo operation on the Hongqi EV platform, the visibility of a target in the point cloud and its geometric representation may vary with observation conditions. To reconstruct such variations in simulation, perception review is modeled as a scenario-search task constrained by real-vehicle records, and its output supports subsequent review of the perception chain.

At time \(t\), the perception-review state \(s_t^{\mathrm{P}}=[\bar d_t^{eo},\bar o_t,\bar h_t,\bar\rho_{20,t},\bar\rho_{h,t},\bar q_t]\) describes the observation condition of the current candidate under constraints retained from the vehicle record. Here, \(\bar d_{t}^{eo}\) is the normalized ego--object distance, \(\bar o_t\) is the target visibility level, \(\bar h_t\) is the target-height condition, \(\bar\rho_{20,t}\) is the near-range point-cloud ratio, \(\bar\rho_{h,t}\) is the low-height point-cloud ratio, and \(\bar q_t\) is the record-quality score. The action \(a_t^{\mathrm{P}}\) adjusts the observation condition of the candidate scenario and takes values from \(a_{t}^{\mathrm{P}}\in\{\Delta d^{-},\Delta d^{+},\Delta o^{-},\Delta o^{+},\Delta h^{-},\Delta h^{+}\}\). After the action is executed, the simulation environment updates the next state according to \(s_{t+1}^{\mathrm{P}}=\mathcal{P}_{\mathrm{P}}(s_{t}^{\mathrm{P}},a_{t}^{\mathrm{P}})\), and the generated trajectory is represented as \(\tau_{\mathrm{P}}=(s_{0}^{\mathrm{P}},a_{0}^{\mathrm{P}},\ldots,s_{T}^{\mathrm{P}})\).

The reward function guides the search toward scenarios with perception-review value while penalizing departures from real-vehicle record constraints, with \(r_{t}^{\mathrm{P}}=\alpha_{1}R_{d}+\alpha_{2}R_{o}+\alpha_{3}R_{h}+\alpha_{4}R_{\rho}-\alpha_{5}R_{\xi}\). In this reward, \(R_d\) measures the review value associated with ego--object distance, \(R_o\) measures the effect of target visibility, \(R_h\) measures the effect of target-height condition, \(R_{\rho}\) constrains consistency with point-cloud statistics from real-vehicle records, and \(R_{\xi}\) is the constraint-violation penalty. The feasible trajectory set of the perception-review task is constrained by the vehicle records, low-speed operation condition, and review requirement, written as \(\Omega_{\mathrm{P}}=\Omega_{\mathrm{P}}^{L}\cap\Omega_{\mathrm{P}}^{S}\cap\Omega_{\mathrm{P}}^{Q}\), where \(\Omega_{\mathrm{P}}^{L}\) constrains point-cloud record consistency, \(\Omega_{\mathrm{P}}^{S}\) constrains the low-speed scenario range, and \(\Omega_{\mathrm{P}}^{Q}\) constrains the reviewability of the candidate scenario. The termination function is defined as \(g_{\mathrm{P}}(s_t)=\mathbb{I}[d_t^{eo}<d_{\min}\vee o_t>o_{\max}\vee t=T_{\max}\vee R_{\xi}>0]\). When \(g_{\mathrm{P}}(s_t)=1\), the current scenario search is terminated. Accordingly, \(\mathcal{C}_{\mathrm P}\) contains the distance-, visibility-, and height-oriented candidates \(c_i^{d}\), \(c_i^{o}\), and \(c_i^{h}\), each mapped from a generating trajectory in \(\Omega_{\mathrm P}\).

\paragraph{Boundary-Response Scenario Generation \(\mathcal{M}_{\mathrm B}\).}

Boundary-response scenario generation is used to review system behavior when low-speed tests approach the safety-margin boundary. During campus-road testing of Apollo on the Hongqi EV platform, changes in the distance and relative motion between the ego vehicle and surrounding objects may alter the system response to the current interaction. To retain reviewable samples under such critical conditions, boundary response is modeled as a scenario-search task constrained by safety margins and road conditions.

At time \(t\), the boundary-response state \(s_t^{\mathrm{B}}=[\bar d_t^{eo},\bar v_t^r,\bar y_t,\bar m_t,\bar u_t]\) relates the generated interaction to the operating limits retained from the low-speed vehicle record. Here, \(\bar d_t^{eo}\) is the normalized ego--object distance, \(\bar v_t^r\) is the relative velocity, \(\bar y_t\) is the lateral offset of the object, \(\bar m_t\) is the safety margin, and \(\bar u_t\) is the system-response descriptor. The margin is computed as \(m_t=d_t^{eo}-d_{\mathrm{safe}}(v_t^e,v_t^r)\), which measures how close the current interaction is to the safety boundary.

Scenario editing is performed by \(a_t^{\mathrm{B}}\in\{\Delta d^{-},\Delta d^{+},\Delta v_r^{-},\Delta v_r^{+},\Delta y^{-},\Delta y^{+}\}\). These operators change the low-speed interaction condition and lead to the next state through \(s_{t+1}^{\mathrm{B}}=\mathcal{P}_{\mathrm{B}}(s_t^{\mathrm{B}},a_t^{\mathrm{B}})\). The generated sequence is recorded as \(\tau_{\mathrm{B}}=(s_0^{\mathrm{B}},a_0^{\mathrm{B}},\ldots,s_T^{\mathrm{B}})\).

The reward is defined as \(r_t^{\mathrm{B}}=\beta_1R_m+\beta_2R_u+\beta_3R_p-\beta_4R_s-\beta_5R_{\xi}\), where \(R_m\) increases the value of samples near the safety-margin boundary, \(R_u\) reflects the observability of the system response, and \(R_p\) reflects the plausibility of the interaction process. The terms \(R_s\) and \(R_{\xi}\) penalize unsafe states and constraint violations. The feasible trajectory set is \(\Omega_{\mathrm{B}}=\Omega_{\mathrm{B}}^{M}\cap\Omega_{\mathrm{B}}^{R}\cap\Omega_{\mathrm{B}}^{V}\), which keeps the search within the admissible margin, road, and low-speed ranges. The search terminates when \(g_{\mathrm{B}}(s_t)=\mathbb{I}[m_t<0\vee d_t^{eo}<d_{\min}\vee t=T_{\max}\vee R_s>0]\) equals one. The retained candidate set is \(\mathcal{C}_{\mathrm{B}}=\{c_i^f,c_i^x,c_i^a\}_{i=1}^{N_{\mathrm{B}}}\), with \(\mathcal{C}_{\mathrm{B}}\subset\Omega_{\mathrm{B}}\).

\paragraph{Anomaly-Review Scenario Generation \(\mathcal{M}_{\mathrm A}\).}

Anomaly-review scenario generation starts from abnormal segments that have already been observed during real-vehicle testing. In the Apollo-on-Hongqi EV records, such segments provide a traceable anchor for reconstructing comparable conditions in simulation. The retained candidates support anomaly replay and follow-up review under controlled test settings.

For anomaly reconstruction, the state is \(s_t^{\mathrm A}=[z_t,\bar d_t^{eo},\bar o_t,\bar v_t^r,\bar\kappa_t,\bar q_t]\), which links the anomaly cue with the interaction and record conditions required for controlled reconstruction. The cue is encoded as \(z_t=\operatorname{Enc}_{\mathrm A}(x_t^{\mathrm{log}},x_t^{\mathrm{view}},x_t^{\mathrm{note}})\), where \(\operatorname{Enc}_{\mathrm A}\) maps the retained log, visualization, and researcher-note evidence into a normalized anomaly-cue score \(z_t\in[0,1]\). The terms \(\bar d_t^{eo}\), \(\bar o_t\), \(\bar v_t^r\), \(\bar\kappa_t\), and \(\bar q_t\) describe the ego--object distance, target visibility, relative motion, test configuration, and record quality after normalization.

Candidate reconstruction uses \(a_t^{\mathrm A}\in\{\Delta d,\Delta o,\Delta v_r,\Delta\kappa\}\), where each \(\Delta\) denotes a predefined discrete edit operator and \(\Delta\kappa\) edits the anomaly-related configuration represented by \(\bar\kappa_t\). The simulator maps the edited candidate to \(s_{t+1}^{\mathrm A}=\mathcal{P}_{\mathrm A}(s_t^{\mathrm A},a_t^{\mathrm A})\), producing \(\tau_{\mathrm A}=(s_0^{\mathrm A},a_0^{\mathrm A},\ldots,s_T^{\mathrm A})\). Similarity to the recorded abnormal segment is evaluated by \(R_{\sigma}=\operatorname{sim}(\phi(c_t),\phi(c^z))\), where \(c_t\) is the current candidate scenario and \(c^z\) is the reference scenario constrained by the anomaly cue. The function \(\phi:\mathcal{X}_{\mathrm A}\rightarrow\mathbb{R}^{p}\) maps an anomaly-review scenario to a \(p\)-dimensional feature representation. For nonzero feature vectors, \(\operatorname{sim}(u,v)=\frac{1}{2}\left(1+\frac{u^{\mathsf T}v}{\lVert u\rVert_2\lVert v\rVert_2}\right)\in[0,1]\).

The reward \(r_t^{\mathrm{A}}=\lambda_1R_{\sigma}+\lambda_2R_{\rho}+\lambda_3R_q-\lambda_4R_s-\lambda_5R_{\xi}\) assigns higher value to candidates that approach the observed anomaly while remaining consistent with the real-vehicle record. The terms \(R_{\sigma}\), \(R_{\rho}\), and \(R_q\) correspond to anomaly similarity, record consistency, and review quality, whereas \(R_s\) and \(R_{\xi}\) penalize unsafe states and constraint violations. The feasible set \(\Omega_{\mathrm{A}}=\Omega_{\mathrm{A}}^{T}\cap\Omega_{\mathrm{A}}^{S}\cap\Omega_{\mathrm{A}}^{Q}\) binds the search to traceable evidence, safe operation, and reviewable outputs. An episode ends when \(g_{\mathrm{A}}(s_t)=\mathbb{I}[R_{\sigma}\ge\eta\vee R_{\xi}>0\vee t=T_{\max}]\) equals one. After filtering, the retained candidate set is \(\mathcal{C}_{\mathrm{A}}=\{c_i^r,c_i^g,c_i^q\}_{i=1}^{N_{\mathrm{A}}}\), with \(\mathcal{C}_{\mathrm{A}}\subset\Omega_{\mathrm{A}}\).

The three task-specific MDPs define the search spaces for review-oriented scenario generation. Each formulation starts from vehicle-record evidence and uses \(\Omega_k\) to constrain simulation-based scenario editing. Table~\ref{tab:rl_scenario_generation_mapping} summarizes the task-specific processing flow for \(\mathcal{M}_{\mathrm P}\), \(\mathcal{M}_{\mathrm B}\), and \(\mathcal{M}_{\mathrm A}\).

\begingroup
\footnotesize
\setlength{\tabcolsep}{3pt}
\setlength{\extrarowheight}{1.5pt}
\setlength{\arrayrulewidth}{0.35pt}
\renewcommand{\arraystretch}{1.25}

\setlength{\LTleft}{0pt plus 1fill}
\setlength{\LTright}{0pt plus 1fill}
\setlength{\LTcapwidth}{\textwidth}

\begin{longtable}{
@{}
>{\centering\arraybackslash}m{0.145\textwidth}
>{\centering\arraybackslash}m{0.205\textwidth}
>{\centering\arraybackslash}m{0.205\textwidth}
>{\centering\arraybackslash}m{0.205\textwidth}
>{\centering\arraybackslash}m{0.145\textwidth}
@{}
}

\caption{Proposed detailed flow of task-specific DQN scenario generation.}
\label{tab:rl_scenario_generation_mapping}\\

\hline
\textbf{Process Node} &
\textbf{Perception Review \(\mathcal{M}_{\mathrm P}\)} &
\textbf{Boundary Response \(\mathcal{M}_{\mathrm B}\)} &
\textbf{Anomaly Review \(\mathcal{M}_{\mathrm A}\)} &
\textbf{Output} \\
\hline
\endfirsthead

\hline
\textbf{Process Node} &
\textbf{Perception Review \(\mathcal{M}_{\mathrm P}\)} &
\textbf{Boundary Response \(\mathcal{M}_{\mathrm B}\)} &
\textbf{Anomaly Review \(\mathcal{M}_{\mathrm A}\)} &
\textbf{Output} \\
\hline
\endhead

\endfoot
\endlastfoot

Evidence localization &
Target visibility; point-cloud change &
Safety margin; boundary approach &
Anomaly timestamp; adjacent records &
Task evidence segment \\
\hline

Segment anchoring &
Point-cloud frame range; observation anchor &
Low-speed interaction range; boundary anchor &
Pre-/post-anomaly window; reconstruction anchor &
Anchored record segment \\
\hline

Condition association &
Sensor input; perception result &
Vehicle state; system response &
Runtime log; test record &
Linked evidence record \\
\hline

Scenario initialization &
Observation condition; initial scenario &
Vehicle--object relation; interaction scenario &
Anomaly condition; reference scenario &
Initial candidate scenario \\
\hline

State assembly &
6-D perception state &
5-D boundary state &
6-D anomaly state &
\(s_0^k\) \\
\hline

Numerical encoding &
Perception-range normalization &
Low-speed-range normalization &
Review-range normalization &
Policy input \\
\hline

Action encoding &
Distance edit; visibility edit; height edit; 6 types &
Distance edit; relative-motion edit; lateral edit; 6 types &
Distance edit; visibility edit; relative-motion edit;
configuration edit; 4 types &
\(\mathcal{A}_k\) \\
\hline

Network mapping &
6-D input; 6 action values &
5-D input; 6 action values &
6-D input; 4 action values &
\(Q_{\theta_k}(s_t^k,\cdot)\) \\
\hline

DQN decision &
Perception-edit action &
Boundary-search action &
Anomaly-reconstruction action &
\(a_t^k\) \\
\hline

Edit execution &
Target observation condition &
Low-speed interaction condition &
Anomaly-related condition &
Edited scenario \\
\hline

Simulation transition &
Perception-state update &
Interaction-state update &
Anomaly-state update &
\(s_{t+1}^k\) \\
\hline

Reward return &
Perception-task reward &
Boundary-task reward &
Anomaly-task reward &
\(r_t^k\) \\
\hline

Feasibility check &
Record consistency; observation boundary &
Safety margin; road boundary &
Evidence traceability; safety condition &
Constraint flag \\
\hline

Termination decision &
Observation out of range; violation; maximum steps &
Safety-margin violation; constraint violation; maximum steps &
Similarity threshold; constraint violation; maximum steps &
\(g_k(s_t^k)\) \\
\hline

Transition recording &
Perception-task identifier &
Boundary-task identifier &
Anomaly-task identifier &
\resizebox{0.98\linewidth}{!}{%
\(\left(
s_t^k;\,
a_t^k;\,
r_t^k;\,
s_{t+1}^k;\,
g_k(s_t^k)
\right)\)} \\
\hline

Trajectory rollout &
Observation-change trajectory &
Interaction-response trajectory &
Anomaly-reconstruction trajectory &
\(\tau_k\) \\
\hline

Candidate validity &
\(\tau_{\mathrm P}\in\Omega_{\mathrm P}\) &
\(\tau_{\mathrm B}\in\Omega_{\mathrm B}\) &
\(\tau_{\mathrm A}\in\Omega_{\mathrm A}\) &
Valid candidate trajectory \\
\hline

Candidate retention &
Distance candidate; visibility candidate; height candidate &
Car-following candidate; crossing candidate; avoidance candidate &
Anomaly-replay candidate; regression-test candidate;
manual-review candidate &
\(\mathcal{C}_k\) \\
\hline

Record association &
Original point-cloud segment; observation anchor &
Original interaction segment; boundary anchor &
Anomaly segment; reconstruction anchor &
Traceable candidate record \\
\hline

Field retention &
Initial condition; edit sequence; terminal state &
Initial condition; interaction sequence; terminal state &
Anomaly cue; edit sequence; terminal state &
Candidate record fields \\
\hline

Review export &
Perception-review material &
Boundary-response material &
Anomaly-review material &
Review package \\
\hline

\end{longtable}
\endgroup

LLM-assisted evidence structuring and reinforcement-learning-based scenario generation preserve their association with the relevant vehicle-record anchors. The LLM output \((o_j,u_j)\) contains evidence-supported fields and unresolved information from Apollo execution records. The RL output keeps each candidate \(c(\tau_i^k)\in\mathcal{C}_k\) associated with its feasible generation trajectory \(\tau_i^k\in\Omega_k\). Linking these outputs to the corresponding Apollo record scope and vehicle operating condition enables researchers to locate the review focus in the software execution path, hardware operating condition, or scenario constraint. The resulting assessment determines the technical direction of the next simulation or HIL test.

The framework retains outputs that satisfy the corresponding review conditions. LLM-derived fields are retained after source verification within the preserved evidence. DQN-generated candidates are retained after their record association and task-constraint satisfaction have been confirmed. Researcher review also accounts for possible implementation faults in reinforcement-learning programs~\cite{song2026rl_bugs}. Each accepted result retains its evidence basis and follow-up test requirement, thereby connecting software-side review, hardware-condition inspection, and scenario reconstruction within the software--hardware collaborative testing process.

\FloatBarrier

\section{Preliminary Result Analysis}
\label{sec:preliminary_results}

The preliminary results are analyzed in relation to the three mechanisms discussed in Sect.~\ref{sec:mechanisms}. The analysis follows the connection among vehicle operation, code reuse, and software-hardware testing, and examines how the available records can be reviewed under the experimental conditions in which they were produced.

\subsection{Multi-Vehicle Collaborative Experiment Evidence}

\begin{figure}[!t]
\centering

\includegraphics[width=0.56\textwidth]{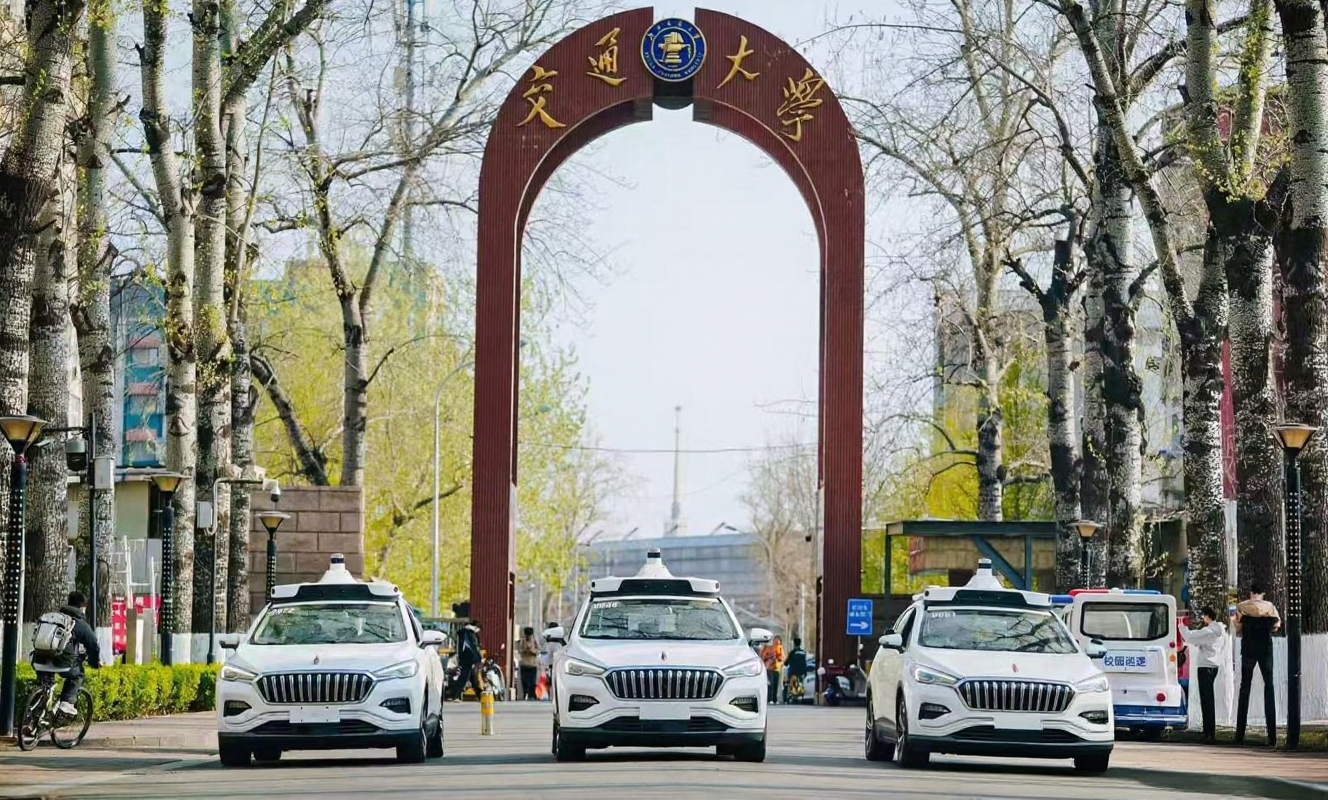}
\par\vspace{1mm}
{\footnotesize (a) Multi-vehicle Hongqi EV experimental platform}

\vspace{3mm}

\includegraphics[
width=0.94\textwidth,
keepaspectratio
]{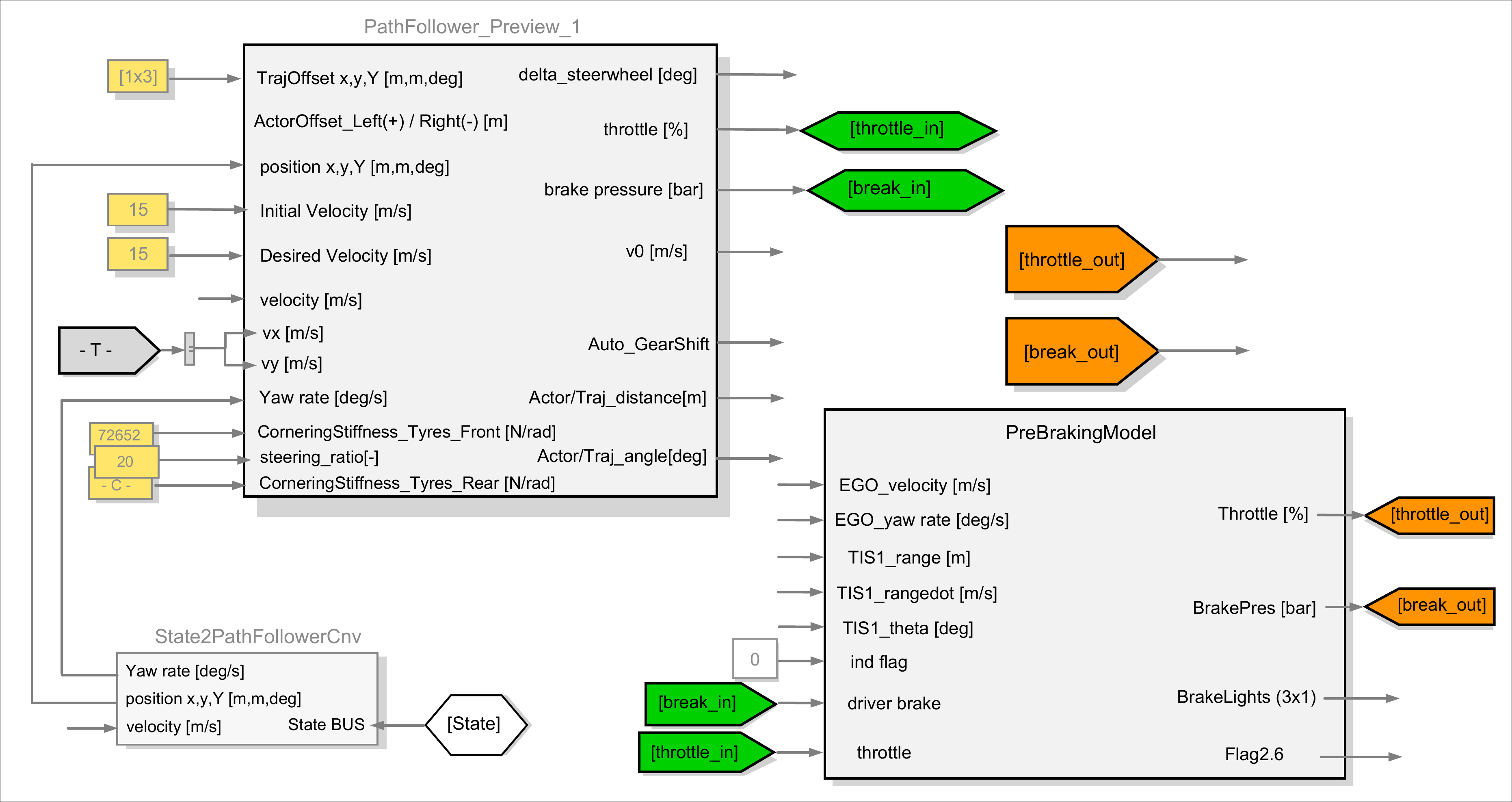}
\par\vspace{1mm}
{\footnotesize (b) Simulink/PreScan control-process reference model}

\caption{Multi-vehicle experimental context and control-process reference:
(a) Hongqi EV platform used in the multi-team environment;
(b) automated-control model used for run-level review.}
\label{fig:multi_vehicle_control_evidence}
\end{figure}

Real-vehicle control research involving multiple vehicles is examined in the Apollo-on-Hongqi EV environment. The research setting uses Hongqi EV test vehicles donated by Baidu, with the Apollo open-source autonomous driving system deployed on the vehicles, as shown in Fig.~\ref{fig:multi_vehicle_control_evidence}(a). Vehicle automation, electrical-system status, and network-security issues are examined within the same real-vehicle operating context. Under comparable vehicle conditions, observations from these directions can be reviewed in relation to the same control process.

The multi-vehicle configuration of the Hongqi EVs gives the study a real-vehicle setting. For control problems such as trajectory tracking and braking response, changes in vehicle state need to be interpreted in relation to the platform condition. When a run is linked with the corresponding vehicle configuration and retained in the experimental record, similar control phenomena can be reviewed under comparable conditions. The comparison can then focus on how observed differences are related to vehicle operation and test settings.

The Simulink/PreScan automated-control model shown in Fig.~\ref{fig:multi_vehicle_control_evidence}(b) provides a reference for the control process. The model describes the relation between control commands and vehicle response in the real-vehicle experiment. For the same control process, control response, chassis response, and abnormal phenomena recorded during operation can be reviewed together. At this stage, the same run offers preliminary material for relating vehicle-automation, electrical-system, and network-security observations.

In this setting, the multi-vehicle environment allows real-vehicle control problems to be observed under comparable vehicle conditions. The automated-control model provides a reference for recording and interpreting the control process, so that control phenomena can be examined together with the vehicle operating context. In the university-industry research setting, such records make shared vehicle use reviewable across teams and offer preliminary material for later cross-vehicle comparison.

\FloatBarrier

\subsection{Code and Experimental-Skill Sharing Evidence}

The code-sharing evidence comes from a patch-based submission related to an extension experiment in the Apollo LiDAR detection chain~\cite{liu2025bft3d}. The submitted code is analyzed together with the configuration switches and verification records that define its operating condition. These records clarify the position of the modification, its activation condition, and the level of evidence available before the branch is used in another vehicle-side task.

Table~\ref{tab:code_reuse_evidence} organizes the submission around its activation condition, verification procedure, and interpretation boundary, so that the branch can be reviewed before real point-cloud evaluation.

\begin{table}[H]
\caption{Review Items for the Code-Submission Case}
\label{tab:code_reuse_evidence}
\centering
\scriptsize
\setlength{\tabcolsep}{4pt}
\renewcommand{\arraystretch}{1.05}
\begin{tabularx}{\textwidth}{@{}>{\raggedright\arraybackslash}m{0.20\textwidth}>{\raggedright\arraybackslash}m{0.43\textwidth}>{\raggedright\arraybackslash}X@{}}
\toprule
Review focus & Recorded information & Review use \\
\midrule

Code location
& Optional proposal-refinement interface inserted between CenterPoint decoding and NMS.
& Identifies where the branch changes the Apollo LiDAR detection chain. \\

Activation path
& Proposal export and refinement are controlled by configuration switches and remain disabled by default.
& Shows whether the extension enters the runtime path under a given configuration. \\

Offline preparation
& The retained preparation process links proposal reading, synthetic-data processing, model preparation, and mode comparison.
& Allows later researchers to reconstruct the preparation and comparison process. \\

Synthetic check
& Four frames with 32 proposals per frame are tested under original, \textit{mock\_zero}, \textit{mock\_delta}, teacher, and student modes.
& Checks whether proposal export, refinement, and write-back paths can run through. \\

Result boundary
& No invalid boxes are reported in the synthetic check; real point-cloud evaluation is not included.
& Limits the conclusion to code-path verification and excludes real detection-performance claims. \\

\bottomrule
\end{tabularx}
\end{table}

The submitted branch is based on the Apollo 11.0.0 master branch. It introduces an optional proposal-refinement extension point between the decoding and non-maximum suppression (NMS) stages of the CenterPoint LiDAR detection pipeline. The new function remains disabled by default, and the original Cyber RT channels and perception message semantics are kept unchanged. These constraints place the modification within a limited LiDAR detection-chain scope and reduce the risk of unexpected influence on the original Apollo runtime behavior.

The synthetic-data check gives a process-level record of the submitted extension. In the reported comparison, the \textit{mock\_zero} mode preserves the original CenterPoint output, which helps check whether the extension point can be inserted without changing the baseline result. The \textit{mock\_delta} mode applies a fixed proposal offset and helps examine whether the delta write-back path is effective. The offline teacher model and the student model further indicate that the runtime refinement interface can be connected with offline model preparation. Since the check uses synthetic data~\cite{qiao2024monosample}, its result is limited to code-path correctness and preparation feasibility; perception effects still need to be examined with real point-cloud data.

The value of this submission lies in its recorded activation path. The patch adds a proposal-refinement interface to the CenterPoint LiDAR detection chain, and the configuration determines whether this interface enters the runtime process. During later use, researchers can restore the submitted branch and check whether proposal output reaches the refinement step in the same way as recorded in the synthetic-data check. This makes it possible to judge the operating condition of the extension before real point-cloud evaluation.

The code submission is recorded together with its configuration condition, verification procedure, and interpretation limit. In the Apollo-on-Hongqi EV research environment, this record allows later teams to distinguish code-path checking from synthetic-data demonstration and real point-cloud evaluation, while keeping the activation condition and verification boundary clear before the branch is adapted to another vehicle or task. Such organization keeps code reuse tied to explicit experimental conditions in later collaborative research.

\subsection{Software-Hardware Collaborative Testing Evidence}

This subsection examines preliminary evidence for the software--hardware collaborative testing mechanism. The real-environment LiDAR record establishes the vehicle-side evidence basis. Controlled perturbation equipment defines traceable test inputs, while board-side inference provides a recorded hardware execution case. A representative LLM-assisted review and a reduced DQN pilot are then used to examine evidence structuring and simulation-based candidate search within the current experimental scope.

\subsubsection{Real-Environment Test Context and LiDAR Evidence}

\begin{figure}[H]
\centering
\begin{minipage}[t]{0.48\textwidth}
\centering
\parbox[c][0.18\textheight][c]{\linewidth}{%
\centering
\includegraphics[width=\linewidth,height=0.18\textheight,keepaspectratio]{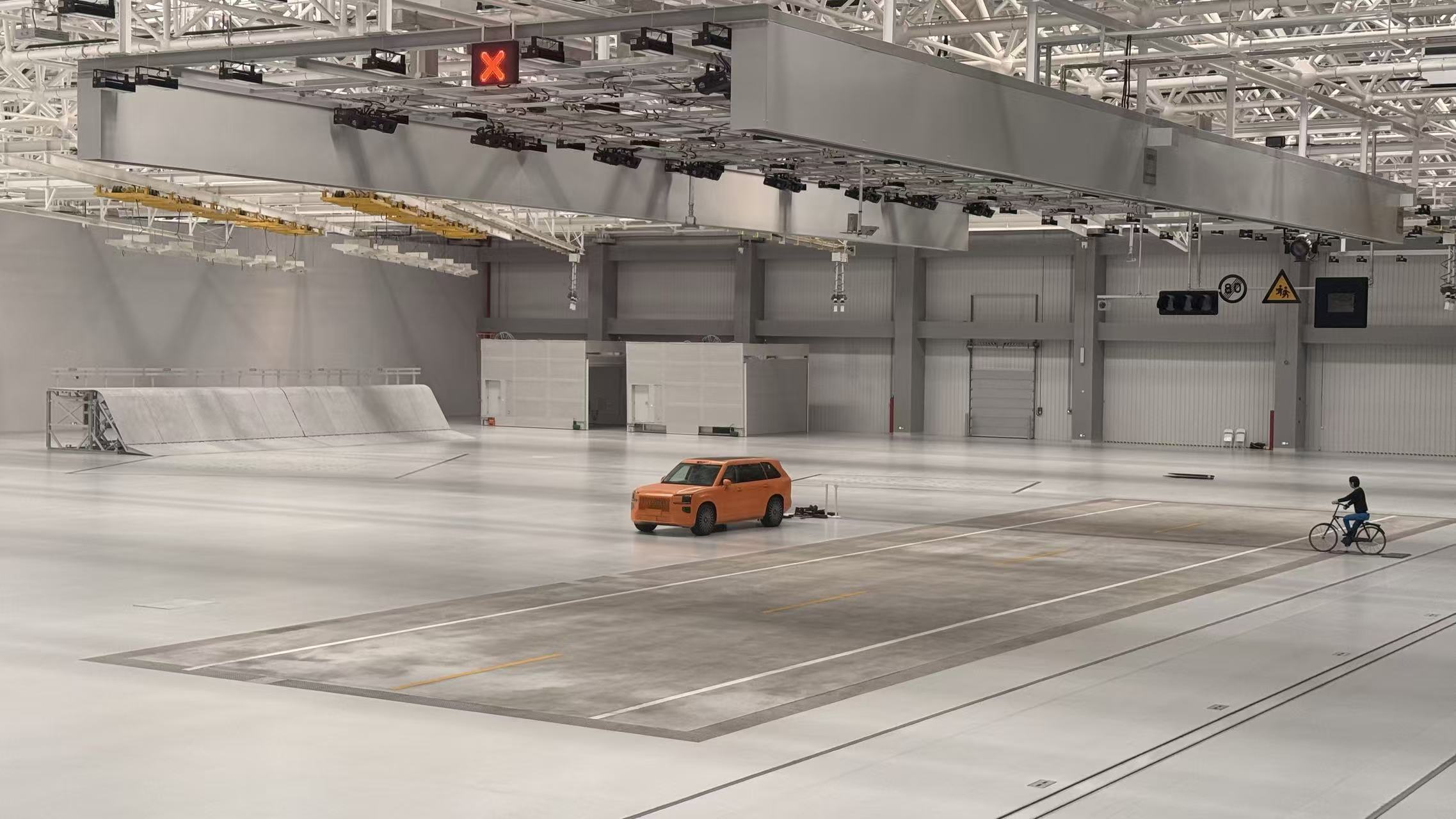}}
\par\vspace{1mm}
{\footnotesize (a) Geely Automobile Research Institute test site}
\end{minipage}
\hfill
\begin{minipage}[t]{0.48\textwidth}
\centering
\parbox[c][0.18\textheight][c]{\linewidth}{%
\centering
\includegraphics[width=\linewidth,height=0.18\textheight,keepaspectratio]{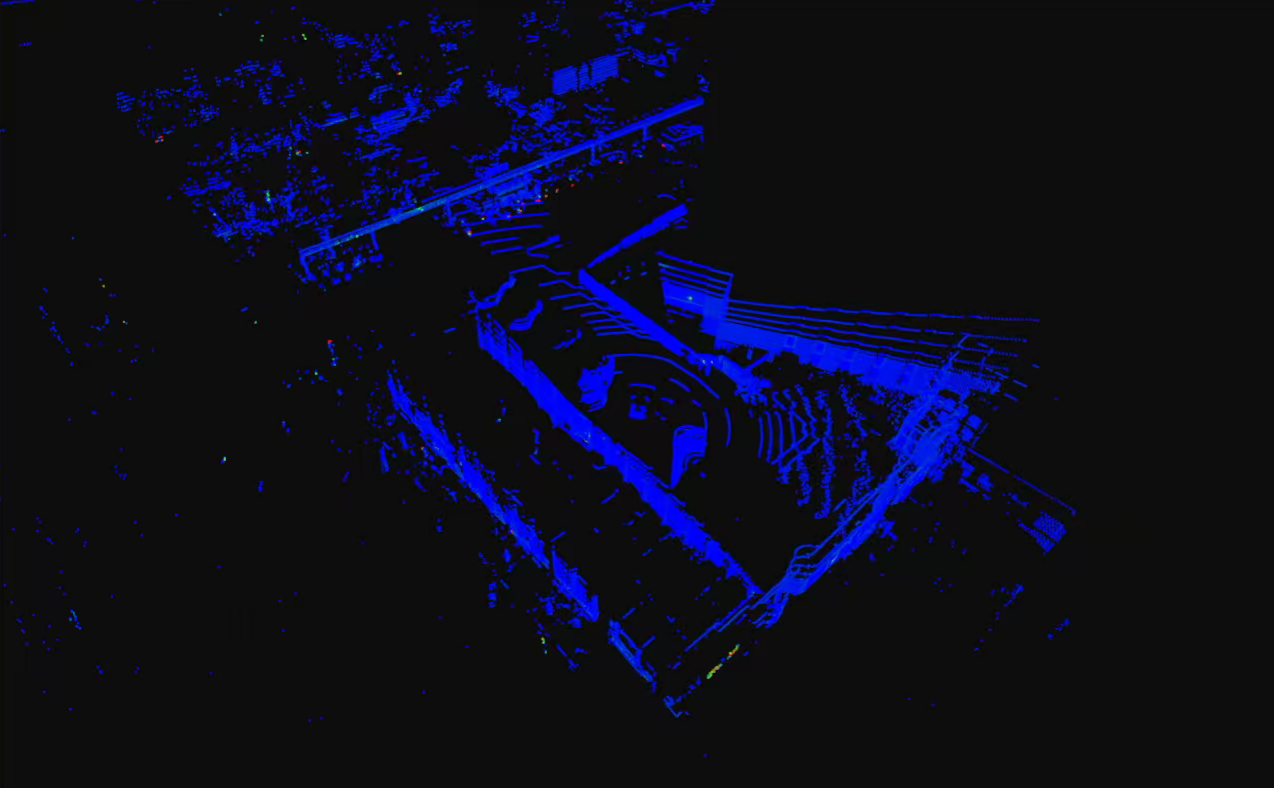}}
\par\vspace{1mm}
{\footnotesize (b) Beijing Jiaotong University campus test site: LiDAR point-cloud view}
\end{minipage}
\caption{Real-environment test contexts used in the Apollo-on-Hongqi EV study.}
\label{fig:real_environment_test_sites}
\end{figure}

Real-environment sensing records are needed to relate test preparation to actual Apollo vehicle operation. Figure~\ref{fig:real_environment_test_sites} distinguishes the two test contexts considered in this subsection. Figure~\ref{fig:real_environment_test_sites}(a) shows the industrial test site at Geely Automobile Research Institute in Zhejiang. Figure~\ref{fig:real_environment_test_sites}(b) presents the LiDAR point-cloud view recorded in the Beijing Jiaotong University campus test environment. The quantitative analysis uses the campus record because its Apollo topic identifier and point-cloud statistics are available for examination.

The campus LiDAR record was obtained from a vehicle-mounted Hesai LiDAR under low-speed driving conditions. The data were recorded through the Apollo Hesai40 point-cloud topic, whose identifier is retained in Table~\ref{tab:bjtu_campus_lidar_record_stats}. The selected scene contains typical campus-road elements and local occlusion, allowing the segment to be examined as a low-speed real-sensor record.

\begin{table}[H]
\centering
\caption{Quantitative Characteristics of the Campus LiDAR Point-Cloud Record}
\label{tab:bjtu_campus_lidar_record_stats}
\scriptsize
\setlength{\tabcolsep}{2.5pt}
\renewcommand{\arraystretch}{1.08}
\begin{tabular}{@{}>{\centering\arraybackslash}m{0.18\textwidth} >{\centering\arraybackslash}m{0.27\textwidth} >{\centering\arraybackslash}m{0.47\textwidth}@{}}
\toprule
Category & Statistical metric & Value \\
\midrule
Data source
& Apollo topic
& \begin{tabular}{@{}c@{}}\texttt{/apollo/sensor/}\\ \texttt{hesai40/PointCloud2}\end{tabular} \\

Temporal feature
& Duration / frames / rate
& 60.035~s / 600 frames / 9.99~Hz \\

Sequence continuity
& Sequence range / discontinuities
& 311--910 / 0 \\

Point-cloud scale
& Total points / mean points per frame
& 41,403,168 / 69,005.28 \\

Point validity
& Valid-point ratio / invalid points
& 100\% / 0 \\

Spatial coverage
& Coordinate range
& \begin{tabular}{@{}c@{}}x: -45.2954--134.6255~m\\ y: -24.3530--158.3474~m\\ z: -5.1968--19.0844~m\end{tabular} \\

Horizontal distance
& Maximum / mean distance
& 177.0068~m / 11.6331~m \\

Range distribution
& Points within 20~m
& 88.57\% \\

Height distribution
& Points from -2~m to 1~m
& 89.65\% \\
\bottomrule
\end{tabular}
\end{table}

The visualization in Fig.~\ref{fig:real_environment_test_sites}(b) is consistent with the statistical concentration of points in near-range and low-height regions. Together with the retained Apollo topic source, the record can be used to examine scenario conditions and real-environment sensing in later analysis.

\subsubsection{Controlled Perturbation Equipment and Test Preparation}
\begin{figure}[H]
\centering
\begin{minipage}[t]{0.24\textwidth}
\centering
\includegraphics[
height=0.20\textheight,
trim=330 0 330 0,
clip,
keepaspectratio
]{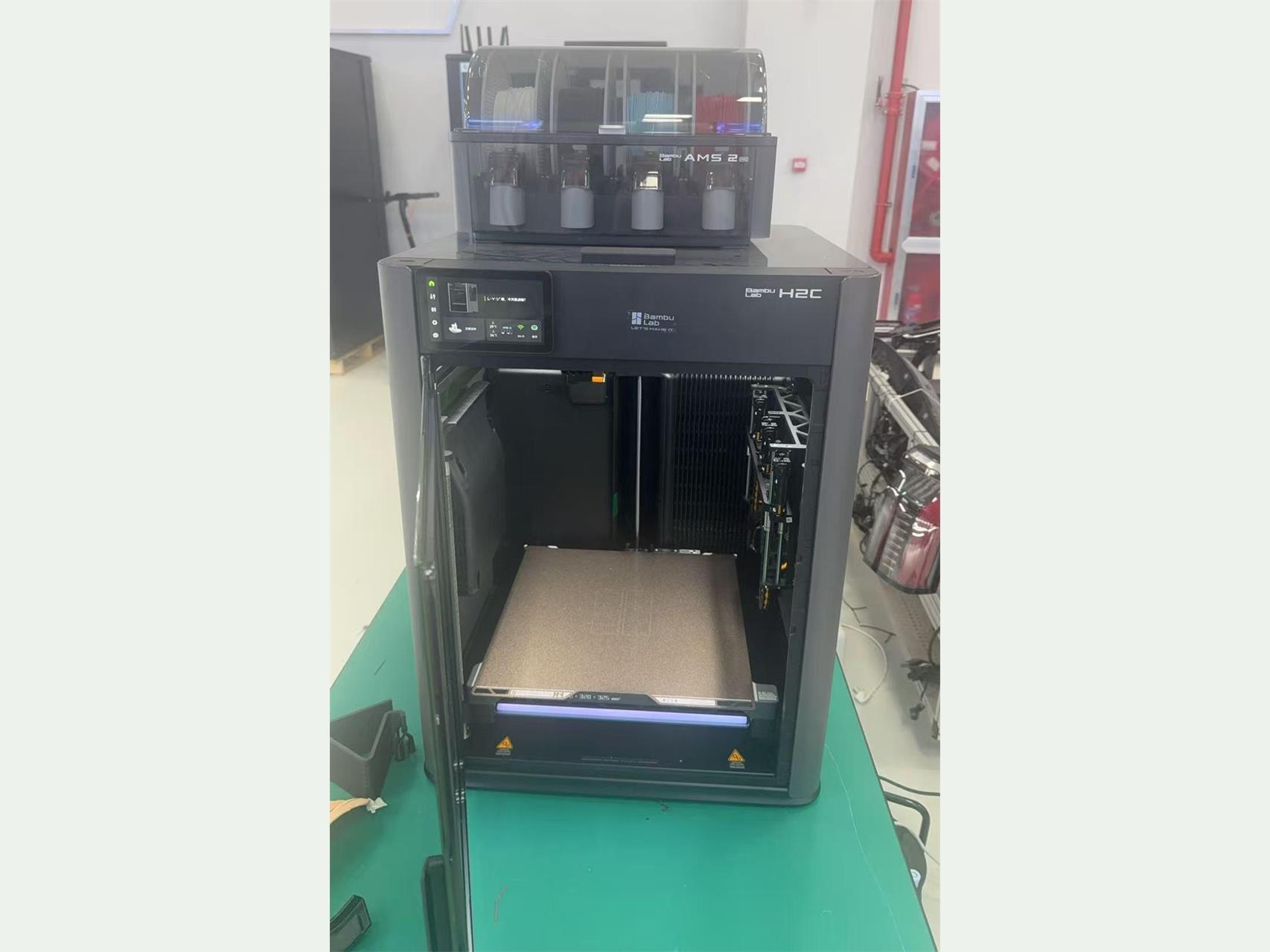}
\par\vspace{1mm}
{\footnotesize (a) 3D-printing platform}
\end{minipage}
\hspace{0.035\textwidth}
\begin{minipage}[t]{0.45\textwidth}
\centering
\includegraphics[
height=0.20\textheight,
keepaspectratio
]{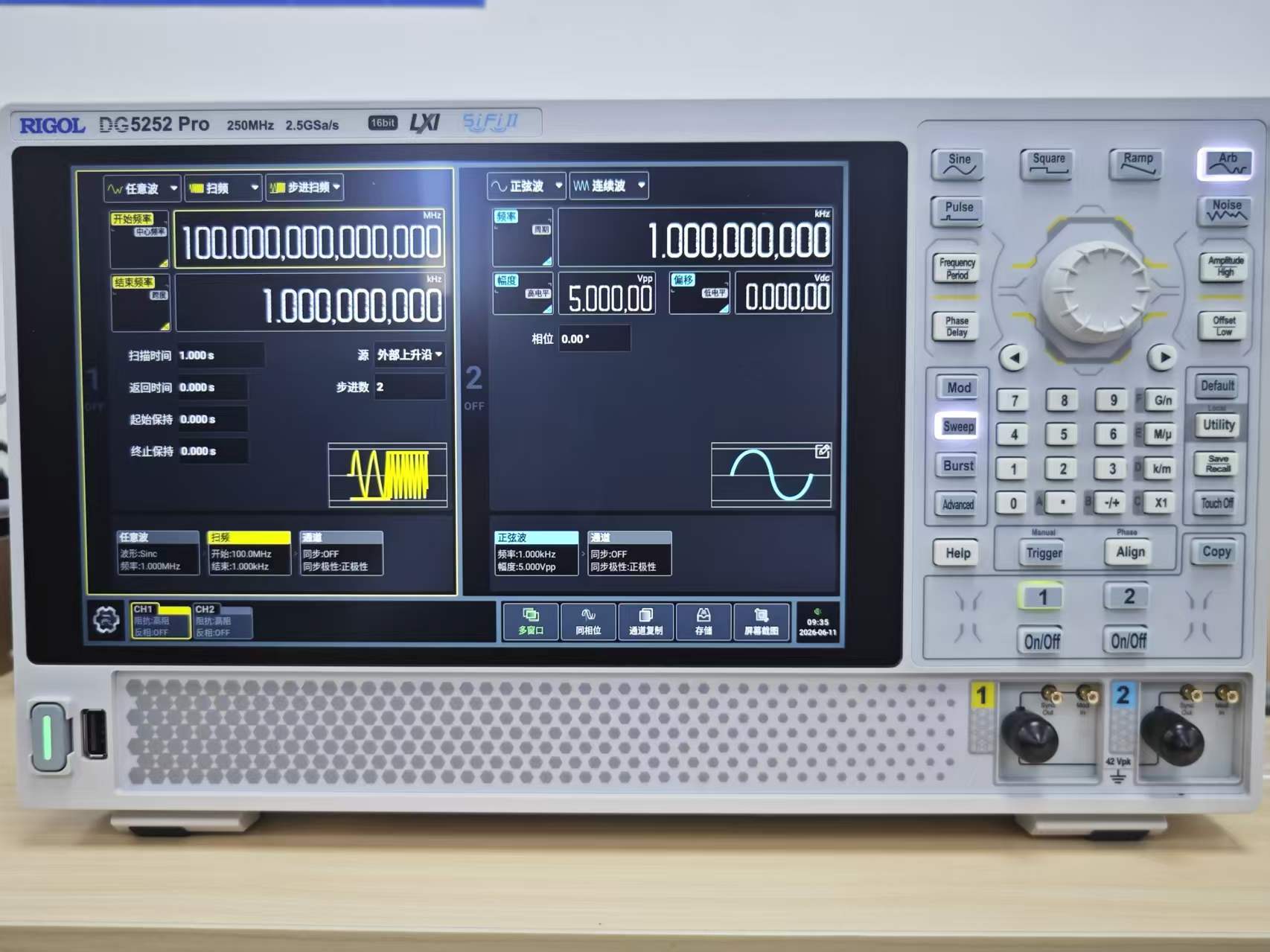}
\par\vspace{1mm}
{\footnotesize (b) Signal function generator}
\end{minipage}
\caption{Adversarial-perturbation analysis equipment for physical-object preparation and configurable signal input.}
\label{fig:adversarial_perturbation_equipment}
\end{figure}

Physical-object preparation and signal-input adjustment are used to define controllable perturbation conditions for the adversarial-perturbation analysis, as shown in Fig.~\ref{fig:adversarial_perturbation_equipment}. The purpose of this test preparation is to prepare object-side and signal-side inputs whose settings can be traced during later vehicle-side review. The 3D-printing platform is used to produce physical targets with specified geometry, surface condition, and mounting configuration. These parameters are retained with the prepared object so that the vehicle-side perception result can be interpreted with reference to a known perturbation source. In this setting, differences in perception outputs can be related to the prepared object condition rather than being treated only as isolated detection results.

The signal function generator provides the corresponding signal-side input condition for perturbation analysis. Its waveform, frequency setting, amplitude range, and sweep configuration are recorded before the vehicle-side or signal-recognition test is performed. These retained settings provide a reference for examining whether changes in sensing outputs or signal-recognition results are associated with the supplied signal condition. The two equipment groups therefore provide controlled experimental inputs from different sides of the software--hardware test process. The physical target preparation supports perception-oriented review, while the signal-input adjustment supports network-security and signal-response review. Together, the retained object and signal settings connect perturbation preparation with subsequent vehicle-side perception analysis and security-related observation.

During the experiment, each perturbation condition was recorded together with the corresponding vehicle-side observation or signal-response output. Only cases with matched perturbation source, equipment setting, and vehicle-side record were retained for later review, so that the observed result could be interpreted within the same experimental scope.

\subsubsection{Edge-Side Execution Evidence}
The MEISHA V100 RISC-V chip board provided by Shenzhen University of Advanced Technology and other collaborators was used for edge-side signal-recognition execution. The board-level platform, shown in Fig.~\ref{fig:meisha_board_evidence}, represents a resource-limited hardware environment for the tested task~\cite{shao2025transient_fault}. The test record reports an 8~mm $\times$ 8~mm bare die, an approximately 3~cm $\times$ 3~cm packaged chip, a 15~cm $\times$ 15~cm board, and 1~MB on-chip SRAM. These values describe the hardware setting under which the board-side inference was recorded.

During the test, the SNN signal-recognition model ran on the MEISHA V100 RISC-V chip board. The record retains the input scale, inference setting, and output statistics of this board-side execution. A loaded weight-file size of 54,994,432 bytes and 5,445,632 IQ data points were recorded, with 265 sample blocks processed over four inference time steps. The main recognized category was the C-V2X PC5 sidelink communication signal, appearing in 254 of the 265 sample blocks, or 95.85\% of the tested blocks. The average confidence recorded for this category was 0.987388. The 1~MB SRAM value describes the on-chip memory setting, while the weight-file size is treated as a loading record. The result is therefore limited to the completion of board-side inference and the generation of structured output under the recorded hardware setting.

\begin{center}
\raisebox{-3mm}{%
\begin{minipage}[c]{0.27\textwidth}
\vspace{0pt}
\centering
\includegraphics[width=\linewidth,keepaspectratio]{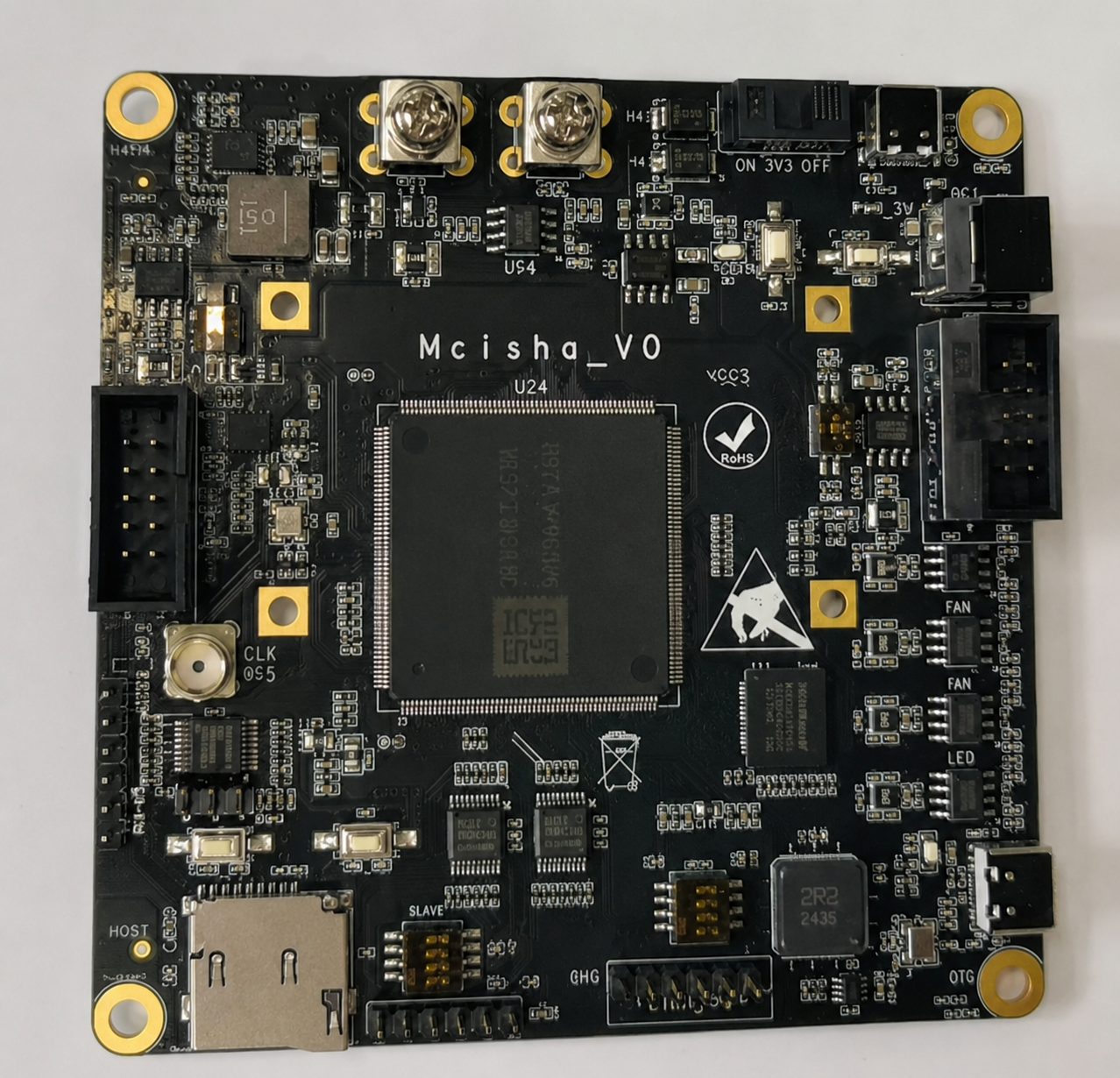}
\captionof{figure}{MEISHA V100 RISC-V chip board.}
\label{fig:meisha_board_evidence}
\end{minipage}%
}
\hspace{0.05\textwidth}
\begin{minipage}[c]{0.66\textwidth}
\vspace{0pt}
\centering
\captionof{table}{Recorded Board-Side SNN Inference on the MEISHA V100 Chip Board}
\label{tab:signal_chip_result}
\scriptsize
\setlength{\tabcolsep}{2.5pt}
\renewcommand{\arraystretch}{1.08}
\begin{tabularx}{\linewidth}{@{}>{\raggedright\arraybackslash}m{0.20\linewidth}>{\raggedright\arraybackslash}m{0.28\linewidth}>{\raggedright\arraybackslash}X@{}}
\toprule
Record focus & Measured item & Recorded value \\
\midrule
Hardware setting
& Chip board; memory
& MEISHA V100 RISC-V; 1~MB SRAM \\

Model loading
& Model; weight size
& SNN signal-recognition; 54,994,432 bytes \\

Input scale
& IQ points; sample blocks
& 5,445,632; 265 \\

Inference setting
& Time steps
& 4 \\

Output record
& Main category
& C-V2X PC5 sidelink signal \\

Output distribution
& Blocks; proportion
& 254 / 265; 95.85\% \\

Confidence record
& Average confidence
& 0.987388 \\
\bottomrule
\end{tabularx}
\end{minipage}
\end{center}

The board-side record keeps the information needed for later examination of the signal-recognition task. When it is considered together with vehicle configuration, Apollo runtime logs, and vehicle-side records, the result can be related to the real-vehicle running state. At the current stage, the evidence describes a recorded execution process and does not amount to a full evaluation of deployment effects.

\subsubsection{Preliminary LLM-Assisted Review and DQN-Guided Scenario Generation}
\paragraph{Representative Runtime-Record Consistency Review.}
This experiment instantiates the runtime-record consistency review task \(\mathcal{L}_{\mathrm R}\) defined in Sect.~\ref{sec:llm_evidence_review}. The retained Apollo topic source and the numerical values in Table~\ref{tab:bjtu_campus_lidar_record_stats} form \(\mathcal{E}_{\mathrm R}\), while \(\Pi_{\mathrm R}\) requires each accepted field to preserve its numerical basis and record scope. The review request \(q_{\mathrm R}^{\mathrm{req}}\) identifies the selected campus LiDAR record and requests a record-grounded review note. DeepSeek-V4-Flash generates the candidate note \(y_{\mathrm R}=\operatorname{LLM}_{\psi_{\mathrm F}}(\operatorname{Prompt}_{\mathrm R}^{\mathrm{gen}}(q_{\mathrm R}^{\mathrm{req}},\mathcal{E}_{\mathrm R},\Pi_{\mathrm R}))\). The candidate note then serves as the statement input to \(\mathcal{L}_{\mathrm R}\) and is decomposed as \(\mathcal{Z}_{\mathrm R}=\operatorname{Split}_{\mathrm R}(y_{\mathrm R},\Pi_{\mathrm R})\). For units \(z_m\in\mathcal{Z}_{\mathrm R}\) requiring closer examination, DeepSeek-V4-Pro performs the second-stage evidence check through \(x_m^{\mathrm R}=\operatorname{Prompt}_{\mathrm R}(z_m,\mathcal{E}_{\mathrm R},\Pi_{\mathrm R})\), with \((\hat{o}_m^{\mathrm R},\hat{u}_m^{\mathrm R})=\operatorname{LLM}_{\psi_{\mathrm P}}(x_m^{\mathrm R})\). Researchers then verify the generated fields through \((o_m^{\mathrm R},u_m^{\mathrm R})=\operatorname{Verify}_{\mathrm R}(\hat{o}_m^{\mathrm R},\hat{u}_m^{\mathrm R};\mathcal{E}_{\mathrm R})\). Fields matched to the table values and Apollo topic source are retained in \(o_m^{\mathrm R}\), while missing, conflicting, or unsupported fields are kept in \(u_m^{\mathrm R}\). The combined Flash--Pro process used fewer than 3k input tokens and fewer than 1k output tokens.

\begin{quote}
\small
\textit{LLM-assisted review note excerpt.}
The retained LiDAR record contains 600 continuous frames over 60.035~s, with no observed sequence discontinuity. The point cloud is dominated by near-range and low-height returns: 88.57\% of the valid points are within 20~m, and 89.65\% fall between -2~m and 1~m in height. These statistics indicate that the segment mainly reflects a low-speed campus-road record. For perception-record review, the Apollo topic source, frame continuity, and spatial distribution were retained as fields requiring comparison with the original record. Detection-performance claims are excluded because the supplied evidence does not contain detection outputs.
\end{quote}

Researcher verification retained the record duration, frame-continuity statement, and spatial-distribution statements in \(o_{\mathrm R}\) because they could be matched to Table~\ref{tab:bjtu_campus_lidar_record_stats} and the corresponding Apollo topic source. Detection-performance statements were excluded because \(\mathcal{E}_{\mathrm R}\) contained no detection output. The unresolved set \(u_{\mathrm R}\) was reserved for fields requiring additional record support. This experiment provides one representative instance of \(\mathcal{L}_{\mathrm R}\), while evaluation of \(\mathcal{L}_{\mathrm E}\) and \(\mathcal{L}_{\mathrm A}\) remains outside the present preliminary analysis.

\paragraph{Reduced Record-Constrained DQN Pilot.}
The same campus LiDAR record was also used in a DQN-guided simulation-scenario generation experiment. This experiment is defined as a reduced mixed pilot and is reported separately from the task-complete evaluations of \(\mathcal{M}_{\mathrm P}\), \(\mathcal{M}_{\mathrm B}\), and \(\mathcal{M}_{\mathrm A}\). The pilot MDP is written as \(\mathcal M_{\mathrm{pilot}}=\langle \mathcal S_{\mathrm{pilot}},\mathcal A_{\mathrm{pilot}},\mathcal P_{\mathrm{pilot}},\mathcal R_{\mathrm{pilot}},\gamma\rangle\). The near-range and low-height point ratios reported in Table~\ref{tab:bjtu_campus_lidar_record_stats} were used to constrain the simulated campus-road condition. The state is written as \(s_t^{\mathrm{pilot}}=[\bar d_t,\bar o_t,\bar v_t^r,\rho_{20},\rho_h]\), where the campus record provides \(\rho_{20}=0.8857\) and \(\rho_h=0.8965\). The action set is \(\mathcal A_{\mathrm{pilot}}=\{\Delta d^{-},\Delta d^{+},\Delta o^{-},\Delta o^{+},\Delta v_r^{-},\Delta v_r^{+}\}\), covering target-distance editing, visibility editing, and relative-motion editing within the reduced search space. The training reward is \(r_t=0.25R_m+0.40R_o+0.25R_c-0.75R_{\xi}\), where \(R_m\) measures the target-distance margin, \(R_o\) evaluates partial occlusion, \(R_c\) preserves consistency with the campus record, and \(R_{\xi}\) penalizes constraint violations. Reviewable states are identified by the independent indicator \(I_{\mathrm{rev}}(s_t)=\mathbb I[8\le d_t\le20 \wedge 0.30\le o_t\le0.85 \wedge 0\le v_t^r\le3.3 \wedge R_{\xi}(s_t)=0]\).

The DQN was trained for 2000 episodes with at most 18 steps per episode and \(\gamma=0.92\). Random sampling was used as the evaluation baseline. The comparison in Table~\ref{tab:dqn_simulation_generation} uses the same evaluation budget for both methods, with five random seeds and 120 evaluation episodes for each seed under the same 18-step budget. An evaluation episode was counted as reviewable when it contained at least one state satisfying \(I_{\mathrm{rev}}(s_t)=1\). The reward values reported in Table~\ref{tab:dqn_simulation_generation} are average rewards over the evaluation steps and are not used as the validity criterion for reviewable episodes. Task-complete evaluations of \(\mathcal{M}_{\mathrm P}\), \(\mathcal{M}_{\mathrm B}\), and \(\mathcal{M}_{\mathrm A}\) remain for subsequent work.

\begin{table}[H]
\centering
\caption{Evaluation of the Reduced DQN Pilot with Independent Reviewability Criteria}
\label{tab:dqn_simulation_generation}
\scriptsize
\setlength{\tabcolsep}{4pt}
\renewcommand{\arraystretch}{1.08}
\begin{tabular}{ccccc}
\toprule
Seed & Random reward & Random reviewable episodes & DQN reward & DQN reviewable episodes \\
\midrule
42 & -0.0832 & 8 (6.67\%) & 0.4897 & 120 (100.00\%) \\
43 & -0.0840 & 6 (5.00\%) & 0.5612 & 120 (100.00\%) \\
44 & -0.1138 & 4 (3.33\%) & 0.5487 & 120 (100.00\%) \\
45 & -0.1164 & 6 (5.00\%) & 0.5127 & 120 (100.00\%) \\
46 & -0.1001 & 7 (5.83\%) & 0.4471 & 119 (99.17\%) \\
\midrule
Mean & -0.0995 & 6.2 (5.17\%) & 0.5119 & 119.8 (99.83\%) \\
Std. & 0.0158 & 1.5 (1.24\%) & 0.0460 & 0.4 (0.37\%) \\
\bottomrule
\end{tabular}
\end{table}

Across five seeds, the DQN policy produced at least one independently reviewable state in an average of 99.83\% of the evaluation episodes, compared with 5.17\% for random sampling. The DQN policy also achieved a higher mean step reward under the same evaluation budget, while reviewability was determined only by \(I_{\mathrm{rev}}(s_t)\). These results provide preliminary evidence for the reduced record-constrained pilot in simulation. Separate experiments are required to evaluate the three task-specific MDP formulations.

The point-cloud record links real-sensor data with the Apollo vehicle recording process through the retained topic source. The continuous sequence and spatial distribution describe the selected campus-road segment under low-speed driving conditions. When considered together with the chip-board execution record and perturbation-preparation resources, this sensing record places hardware-side execution, controlled test preparation, and real-vehicle feedback within the same research context. The current evidence remains preliminary; its use in broader collaborative research should be supported by additional vehicle records, controlled experiments, and comparative review.
\FloatBarrier

\section{Discussion and Conclusion}
\label{sec:discussion_conclusion}

This paper examines the Apollo-on-Hongqi EV real-vehicle research environment and discusses how an open-source autonomous driving system can support multidisciplinary HIL studies. The work addresses a practical challenge in shared vehicle experiments, where vehicle runs, code changes, hardware tests, and field data are often produced by different teams at different stages. When these materials are separated from their operating conditions, later review and reuse become difficult. By linking vehicle operation with Apollo runtime records and experimental records, the study provides a way to trace real-vehicle tests and compare results across tasks.

The analysis suggests that the value of the Apollo-on-Hongqi EV platform lies in keeping each experiment connected with its original operating conditions. This connection reduces the separation among shared vehicle use, repository-based code work, and hardware-side testing feedback. It also allows outputs from one stage to be checked against records from another stage during later review.

The result analysis examines the proposed framework through three groups of preliminary evidence. The multi-vehicle evidence relates shared vehicle use to retained operating conditions. The code-submission case connects an Apollo modification with its activation condition and reuse boundary. The board-side inference record documents hardware execution under a specified configuration, while the perturbation equipment defines controllable test inputs for later review. The campus LiDAR record provides the evidence source for a representative \(\mathcal L_{\mathrm R}\) review and a reduced DQN simulation pilot. Together, these results show how vehicle operation, code reuse, hardware execution, test preparation, and real-sensor records can be organized within the same software--hardware collaborative review process.

The main contribution of this paper is the organization of a traceable real-vehicle research process for Apollo-on-Hongqi EV. The Baidu-donated vehicle platform provides the experimental basis for retaining vehicle-side records, software-adaptation evidence, and testing-preparation materials within a reviewable university research setting. The current evidence remains limited in scale. The representative \(\mathcal L_{\mathrm R}\) case shows how a domestically developed large language model can support evidence-constrained organization of Apollo records, with DeepSeek-V4-Flash generating a schema-constrained candidate review record within the reported token budget and DeepSeek-V4-Pro supporting a second-stage consistency check under explicit evidence boundaries. The reinforcement-learning evaluation uses a reduced search space with random sampling as its baseline, and task-complete evaluation of the remaining LLM tasks and the three MDP formulations requires additional experiments. Future work will examine repeated vehicle runs, controlled perturbation trials, stronger search baselines, and broader HIL comparisons.
\section*{Funding}

This work is supported by the National Natural Science Foundation of China under Grant No.~62372021.

\section*{Acknowledgements}

The authors gratefully acknowledge the Data Security Department, Baidu, Inc., for its support with the Apollo-on-Hongqi EV research environment, technical discussions, and experimental conditions. The authors also thank the collaborating academic team from Chang'an University for discussions on Apollo-based vehicle experimentation and cross-team review, and the Geely Automobile Research Institute, Zhejiang, China, for support with vehicle testing conditions and research exchange.


\begin{thebibliography}{99}

\bibitem{huai2023doppeltest}
Y. Huai, Y. Chen, S. Almanee, T. Ngo, X. Liao, Z. Wan, Q. A. Chen, and J. Garcia.
\newblock Doppelg{\"a}nger test generation for revealing bugs in autonomous driving software.
\newblock In \emph{Proc. 2023 IEEE/ACM 45th International Conference on Software Engineering (ICSE)}, 2023, pp. 2591--2603. doi: 10.1109/ICSE48619.2023.00216.

\bibitem{ding2023safety_critical_scenario_survey}
W. Ding, C. Xu, M. Arief, H. Lin, B. Li, and D. Zhao.
\newblock A survey on safety-critical driving scenario generation: A methodological perspective.
\newblock \emph{IEEE Transactions on Intelligent Transportation Systems}, vol. 24, no. 7, pp. 6971--6988, 2023. doi: 10.1109/TITS.2023.3259322.

\bibitem{song2024scenario_based_testing_path}
Q. Song, E. Engstr\"om, and P. Runeson.
\newblock An empirically grounded path forward for scenario-based testing of autonomous driving systems.
\newblock In \emph{Companion Proc. 32nd ACM International Conference on the Foundations of Software Engineering (FSE Companion)}, 2024, pp. 232--243. doi: 10.1145/3663529.3663843.

\bibitem{chen2024misconfiguration_ads}
Y. Chen, Y. Huai, S. Li, C. Hong, and J. Garcia.
\newblock Misconfiguration software testing for failure emergence in autonomous driving systems.
\newblock \emph{Proceedings of the ACM on Software Engineering}, vol. 1, no. FSE, Art. no. 85, 2024. doi: 10.1145/3660792.

\bibitem{chen2025bug_fix_patterns_ads}
Y. Chen, Y. Huai, Y. He, S. Li, C. Hong, Q. A. Chen, and J. Garcia.
\newblock A comprehensive study of bug-fix patterns in autonomous driving systems.
\newblock \emph{Proceedings of the ACM on Software Engineering}, vol. 2, no. FSE, Art. no. FSE018, 2025. doi: 10.1145/3715733.

\bibitem{tu2025trafficcomposer}
Z. Tu, L. Niu, W. Fan, and T. Zhang.
\newblock Multi-modal traffic scenario generation for autonomous driving system testing.
\newblock \emph{Proceedings of the ACM on Software Engineering}, vol. 2, no. FSE, Art. no. FSE078, 2025. doi: 10.1145/3729348.

\bibitem{song2026rapid_review_ads_testing}
Q. Song, A. Nouri, H. Sivencrona, and F. Sarro.
\newblock From research to practice: An interactive rapid review of autonomous driving system testing in industry.
\newblock arXiv:2605.00531, 2026. doi: 10.48550/arXiv.2605.00531.

\bibitem{zhao2024ad_frameworks_simulators}
H. Zhao, M. Meng, X. Li, J. Xu, L. Li, and S. Galland.
\newblock A survey of autonomous driving frameworks and simulators.
\newblock \emph{Advanced Engineering Informatics}, vol. 62, Art. no. 102850, 2024. doi: 10.1016/j.aei.2024.102850.

\bibitem{aliane2025oss_ads_survey}
N. Aliane.
\newblock A survey of open-source autonomous driving systems and their impact on research.
\newblock \emph{Information}, vol. 16, no. 4, Art. no. 317, 2025. doi: 10.3390/info16040317.

\bibitem{jung2025autoware_apollo_compare}
H.-Y. Jung, D.-H. Paek, and S.-H. Kong.
\newblock Open-source autonomous driving software platforms: Comparison of Autoware and Apollo.
\newblock arXiv:2501.18942, 2025. doi: 10.48550/arXiv.2501.18942.

\bibitem{apollo_open_autonomous_driving_platform}
Baidu Apollo.
\newblock Apollo Open Autonomous Driving Platform.
\newblock GitHub repository, 2026. [Online]. Available: \url{https://github.com/ApolloAuto/apollo}. Accessed: Jun. 5, 2026.

\bibitem{apollo_cyber_rt_framework}
Baidu Apollo.
\newblock Apollo Cyber RT framework.
\newblock [Online]. Available: \url{https://developer.apollo.auto/cyber.html}. Accessed: Jun. 5, 2026.

\bibitem{apollo_cyber_developer_tools}
Baidu Apollo.
\newblock Apollo Cyber RT Developer Tools.
\newblock [Online]. Available: \url{https://apollo.baidu.com/docs/apollo/9.x/md_cyber_2docs_2cyber__developer__tools.html}. Accessed: Jun. 5, 2026.

\bibitem{guo2024autoware_dataspeed_dbw}
H. Guo, J. Li, N. K. Saravanan, J. Wishart, and J. Zhao.
\newblock Developing an automated vehicle research platform by integrating Autoware with the DataSpeed drive-by-wire system.
\newblock SAE Technical Paper 2024-01-1981, 2024. doi: 10.4271/2024-01-1981.

\bibitem{wuersching2024cr2aw}
G. W\"ursching, T. Mascetta, Y. Lin, and M. Althoff.
\newblock Simplifying sim-to-real transfer in autonomous driving: Coupling Autoware with the CommonRoad motion planning framework.
\newblock In \emph{Proc. 35th IEEE Intelligent Vehicles Symposium (IV)}, 2024, pp. 1462--1469. doi: 10.1109/IV55156.2024.10588748.

\bibitem{kaljavesi2024carla_autoware_bridge}
G. Kaljavesi, T. Kerbl, T. Betz, K. Mitkovskii, and F. Diermeyer.
\newblock CARLA-Autoware-Bridge: Facilitating autonomous driving research with a unified framework for simulation and module development.
\newblock In \emph{Proc. 35th IEEE Intelligent Vehicles Symposium (IV)}, 2024, pp. 224--229. doi: 10.1109/IV55156.2024.10588623.

\bibitem{gulzar2024roundabouts_unprotected_turns}
M. Gulzar, Y. Muhammad, and N. Muhammad.
\newblock Navigating roundabouts and unprotected turns in autonomous driving.
\newblock \emph{IEEE Transactions on Field Robotics}, vol. 1, pp. 27--46, 2024. doi: 10.1109/TFR.2024.3421389.

\bibitem{lucchetti2023resilient_apollo}
F. Lucchetti, R. Graczyk, and M. V\"olp.
\newblock Toward resilient autonomous driving---An experience report on integrating resilience mechanisms into the Apollo autonomous driving software stack.
\newblock \emph{Frontiers in Computer Science}, vol. 5, Art. no. 1125055, 2023. doi: 10.3389/fcomp.2023.1125055.

\bibitem{oh2024ad_vils}
T. Oh, Y. Ha, D. Yoo, and J. Yoo.
\newblock AD-VILS: Implementation and reliability validation of vehicle-in-the-loop simulation platform for evaluating autonomous driving systems.
\newblock \emph{IEEE Access}, vol. 12, pp. 164190--164209, 2024. doi: 10.1109/ACCESS.2024.3492162.

\bibitem{oh2024xil_reliability}
T. Oh, S. Cho, and J. Yoo.
\newblock A reliability evaluation methodology for X-in-the-loop simulation in autonomous vehicle systems.
\newblock \emph{IEEE Access}, vol. 12, pp. 193622--193640, 2024. doi: 10.1109/ACCESS.2024.3519713.

\bibitem{coppola2025ccam_vil}
A. Coppola, A. Mungiello, G. Pane, A. Petrillo, and S. Santini.
\newblock On the virtual testing of ADAS in CCAM environment via vehicle-in-the-loop framework.
\newblock \emph{IFAC-PapersOnLine}, vol. 59, no. 5, pp. 151--156, 2025. doi: 10.1016/j.ifacol.2025.07.097.

\bibitem{lu2023deepscenario}
C. Lu, T. Yue, and S. Ali.
\newblock DeepScenario: An open driving scenario dataset for autonomous driving system testing.
\newblock Zenodo, Mar. 2023. doi: 10.5281/zenodo.7714194.

\bibitem{zhang2024chatscene}
J. Zhang, C. Xu, and B. Li.
\newblock ChatScene: Knowledge-enabled safety-critical scenario generation for autonomous vehicles.
\newblock In \emph{Proc. IEEE/CVF Conf. Computer Vision and Pattern Recognition (CVPR)}, 2024, pp. 15459--15469. doi: 10.1109/CVPR52733.2024.01464.

\bibitem{odu2025llm_safety_case_apollo}
O. Odu, A. B. Belle, and S. Wang.
\newblock LLM-based safety case generation for Baidu Apollo: Are we there yet?
\newblock In \emph{Proc. IEEE/ACM 4th International Conf. AI Engineering--Software Engineering for AI (CAIN)}, 2025, pp. 222--233. doi: 10.1109/CAIN66642.2025.00033.

\bibitem{liu2023rl_scenario_editing}
H. Liu, L. Zhang, S. K. S. Hari, and J. Zhao.
\newblock Safety-critical scenario generation via reinforcement learning based editing.
\newblock arXiv:2306.14131, 2023. doi: 10.48550/arXiv.2306.14131.

\bibitem{yang2025denserl_testing}
J. Yang, R. Bai, H. Ji, Y. Zhang, J. Hu, and S. Feng.
\newblock Adaptive testing environment generation for connected and automated vehicles with dense reinforcement learning.
\newblock \emph{IEEE Transactions on Intelligent Transportation Systems}, vol. 26, no. 4, pp. 5135--5145, 2025. doi: 10.1109/TITS.2025.3535866.

\bibitem{li2023robust_rl}
Y. Li, Y. Tian, E. Tong, W. Niu, and J. Liu.
\newblock Robust reinforcement learning via progressive task sequence.
\newblock In \emph{Proc. 32nd International Joint Conference on Artificial Intelligence (IJCAI)}, 2023, pp. 455--463. doi: 10.24963/ijcai.2023/51

\bibitem{li2022multiagent_signal}
Y. Li, W. Niu, Y. Tian, T. Chen, Z. Xie, Y. Wu, Y. Xiang, E. Tong, T. Baker, and J. Liu.
\newblock Multiagent reinforcement learning-based signal planning for resisting congestion attack in green transportation.
\newblock \emph{IEEE Transactions on Green Communications and Networking}, vol. 6, no. 3, pp. 1448--1458, 2022. doi: 10.1109/TGCN.2022.3162649.

\bibitem{song2026rl_bugs}
J. Song, Y. Li, Y. Tian, H. Ma, H. Li, J. Zuo, J. Liu, and W. Niu.
\newblock Investigating the bugs in reinforcement learning programs: Insights from Stack Overflow and GitHub.
\newblock \emph{Automated Software Engineering}, vol. 33, no. 1, Art. no. 9, 2026. doi: 10.1007/s10515-025-00555-z.

\bibitem{yang2026task_scheduling}
L. Yang, M. Yuan, Y. Liu, X. Qu, Z. Hu, Z. Zhang, X. Zhao, and S. Fang.
\newblock Optimization of task scheduling and resource allocation for autonomous vehicle testing in vehicle-road-cloud collaborative systems
\newblock \emph{Expert Systems With Applications}, vol. 299, Art. no. 129943, 2026. doi: 10.1016/j.eswa.2025.129943.

\bibitem{yang2025critical_pedestrian}
L. Yang, S. Liu, S. Feng, H. Wang, X. Zhao, G. Qu, and S. Fang.
\newblock Generation of critical pedestrian scenarios for autonomous vehicle testing.
\newblock \emph{Accident Analysis \& Prevention}, vol. 214, Art. no. 107962, 2025. doi: 10.1016/j.aap.2025.107962.

\bibitem{liu2025bft3d}
B. Liu and Y. Wu.
\newblock BFT3D: A robust BEV feature transformation module for multisensor 3-D object detection.
\newblock \emph{IEEE Sensors Journal}, vol. 25, no. 15, pp. 30175–30185, 2025. doi: 10.1109/JSEN.2025.3580425.

\bibitem{qiao2024monosample}
J. Qiao, B. Liu, J. Yang, B. Wang, S. Xiu, X. Du, and X. Nie.
\newblock MonoSample: Synthetic 3D data augmentation method in monocular 3D object detection.
\newblock \emph{IEEE Robotics and Automation Letters}, vol. 9, no. 8, pp. 7326--7332, 2024. doi: 10.1109/LRA.2024.3414272.

\bibitem{wang2025mhtraj}
J. Chen, C. Jia, W. Xie, D. Zhu, X. Shao and Z. Wang.
\newblock MHTraj: A multi-domain hybrid graph neural network with causal-spatial modeling for multi-agent trajectory prediction.
\newblock \emph{IEEE Transactions on Network Science and Engineering}, 2025. doi: 10.1109/TNSE.2025.3602212.

\bibitem{wang2026maafooc}
J. Chen, M. Chen, D. Zhu, Y. Mao, W. Xie, and Z. Wang.
\newblock MAAFOcc: Multimodal adaptive asymmetric fusion based occupancy prediction.
\newblock \emph{Knowledge-Based Systems}, 2026.

\bibitem{xiao2025offroute}
X. Xiao, W. Xing, R. Gao, and M. Wu.
\newblock No longer getting lost on fork road: Vehicle off-route detection via multi-sensor integration.
\newblock \emph{IEEE Transactions on Intelligent Vehicles}, vol. 10, no. 1, pp. 720--733, 2025. doi: 10.1109/TIV.2024.3417411.

\bibitem{miao2025adaptive_sensor_attack}
Z. Miao, C. Shao, H. Li, Y. Cui, and Z. Tang.
\newblock Adaptive sensor attack detection and defense framework for autonomous vehicles based on density.
\newblock \emph{Computers \& Security}, vol. 148, Art. no. 104149, 2025. doi: 10.1016/j.cose.2024.104149.

\bibitem{shao2025transient_fault}
C. Shao, X. Luo, H. Li, and Z. Tang.
\newblock Transient fault detection and failure effect analysis based on design for test and fault tree analysis for automotive chips.
\newblock \emph{IEEE Transactions on Circuits and Systems I: Regular Papers}, IEEE TCAS-I, vol. 72, no. 12, pp. 7808–7821, 2025. doi: 10.1109/TCSI.2025.3570877.

\end{thebibliography}
\end{document}